\documentclass[a4paper]{article}

\usepackage{a4wide}
\usepackage[utf8]{inputenc} 

\usepackage{comment}
\usepackage{graphicx}
\usepackage{url}
\usepackage{tabularx}
\usepackage{hyperref}
\usepackage{graphicx}
\usepackage{todonotes}
\usepackage{multirow}
\usepackage{amsmath}
\usepackage{authblk}
\usepackage{float}

\begin{document}

\title{Public Discourse about COVID-19 Vaccinations: A Computational Analysis of the Relationship between Public Concerns and Policies}

\author[1,2]{Katarina Boland\thanks{katarina.boland@hhu.de}}   
\author[3]{Christopher Starke\thanks{christopher.starke@uva.nl}} %
\author[2]{Felix Bensmann\thanks{felix.bensmann@gesis.org}} 
\author[1]{Frank Marcinkowski\thanks{Frank.Marcinkowski@phil.hhu.de}} %
\author[1,2]{Stefan Dietze\thanks{stefan.dietze@hhu.de}} %

\affil[1]{Heinrich-Heine-University D\"{u}sseldorf, Universit\"{a}tsstrasse 1, 40225 D\"{u}sseldorf, Germany}

\affil[2] {GESIS - Leibniz Institute for the Social Sciences, Unter Sachsenhausen 6-8, 50667, Cologne, Germany}

\affil[3] {University of Amsterdam, Nieuwe Achtergracht 166, 1018 WV, Amsterdam, The Netherlands}

\date{01.06.2023}

\maketitle

\begin{abstract} 
Societies worldwide have witnessed growing rifts separating advocates and opponents of vaccinations and other COVID-19 countermeasures. With the rollout of vaccination campaigns, German-speaking regions exhibited much lower vaccination uptake than other European regions. While Austria, Germany, and Switzerland (the DACH region) caught up over time, it remains unclear which factors contributed to these changes. Scrutinizing public discourses can help shed light on the intricacies of vaccine hesitancy and inform policy-makers tasked with making far-reaching decisions: policies need to effectively curb the spread of the virus while respecting fundamental civic liberties and minimizing undesired consequences. 
This study draws on Twitter data to analyze the topics prevalent in the public discourse. It further maps the topics to different phases of the pandemic and policy changes to identify potential drivers of change in public attention.
We use a hybrid pipeline to detect and analyze vaccination-related tweets using topic modeling, sentiment analysis, and a minimum of social scientific domain knowledge to analyze the discourse about vaccinations in the light of the COVID-19 pandemic in the DACH region.
We show that skepticism regarding the severity of the COVID-19 virus and towards efficacy and safety of vaccines were among the prevalent topics in the discourse on Twitter but that the most attention was given to debating the theme of freedom and civic liberties. Especially during later phases of the pandemic, when implemented policies restricted the freedom of unvaccinated citizens, increased vaccination uptake could be observed. At the same time, increasingly negative and polarized sentiments emerge in the discourse. This suggests that these policies might have effectively attenuated vaccination hesitancy but were not successfully dispersing citizens' doubts and concerns.
\end{abstract}

\section{Introduction}
The outbreak of the COVID-19 pandemic has fundamentally disrupted societies around the world. To protect the general public and particularly vulnerable groups, governments introduced policies to limit the spread of this infectious disease. These policies included mask mandates, closing schools and the retail sector, curfews, strict lockdowns, and contact restrictions. While many of these measures are assumed to have successfully slowed the spread of the pandemic and contributed to saving lives, they also had detrimental side effects. For instance, a stalled economy led to widespread unemployment \cite{blustein2020work}, lockdown policies strained people's mental health, and were accompanied by an increase in domestic violence \cite{piquero2021domestic}. 
This balancing act forced governments to make difficult trade-off decisions in designing policies with maximum effectiveness and minimal invasiveness. A prime example of this tightrope walk is the COVID-19 vaccination strategies. On the one hand, empirical evidence suggested that widespread vaccination uptake ranked among the most effective means to protect the population from getting infected or hospitalized \cite{andrews2022covid}. Thus, encouraging vaccination uptake was primarily seen as a promising strategy to speed up the reopening of society. Yet, on the other hand, invasive policies to promote or even coerce widespread vaccine uptake, such as vaccine mandates or limiting rights for unvaccinated citizens, mark severe encroachment of civil liberties and spurred a significant backlash among the citizenry \cite{bardosh2022unintended}. For instance, invasive policies increased polarization among citizens, undermining social peace and democracy \cite{jiang2021polarization}. 
Therefore, considering citizens'concerns is crucial when designing adequate vaccination policies and minimizing negative side-effects on society. 
In particular, the German-speaking so-called DACH region (Germany, Austria, Switzerland) exhibited higher vaccine hesitancy than other European countries at the start of the pandemic \cite{DessonEtAl2022}. 

While polling is currently the dominant strategy for governments to retrieve citizens' attitudes, opinions, and behaviors \cite{rothmayr2002government}, online public discourses are gaining ground as an additional source of information \cite{CeronAndNegri2015, ceronAndNegri2016, rubinstein2016ten}. Appropriate data are often retrieved from social networking sites (SNS) such as Twitter, Facebook, or YouTube and then analyzed using Natural Language Processing (NLP) methods. While this type of data has been criticized for its lack of representativeness and oversampling of younger, privileged, and tech-savvy people \cite{hargittai2020potential}, they also offer several distinct characteristics that distinguish them from traditional survey data. First, discourse data are highly dynamic, cheap, and contain information about billions of users \cite{ceron2014every}. Second, they can be analyzed ex-post, thereby shedding light on longitudinal fluctuations of public opinion about real-life events \cite{conradEtAl2021}. Third, they allow for a more nuanced understanding of the ambivalence in public opinion \cite{foad2021limitations} by providing insights into the discursive construction of contentious issues \cite{burnap2015detecting}.

 Hence, many scholars have argued that survey and discourse data can complement each other \cite{BuntainEtAl2016,diazEtAl2016,HillEtAl2020,stierEtAl2020}, particularly in highly dynamic policy contexts such as the COVID-19 pandemic. Indeed, attitudes drawn from surveys and online discourses show similar long-term trends, with discourse data sentiments being more prone to short-term fluctuations \cite{pasekEtAl2020} and polarized opinions about COVID-19 measures being more pronounced in survey data \cite{reiter-haasEtAl2022}. 

Especially in times of crisis, continuously monitoring how public opinion changes is crucial for policy-makers to make informed decisions. For instance, discourse data can help policy-makers assess salient topics and concerns that may lead to vaccine hesitancy, evaluate the level of polarization of suggested policies, or retrieve information about side effects in real-time. 

Our study investigates the potential of online discourse data to inform policy-making about the vaccination strategy in the DACH region. We combined automated methods with manual analyses to harvest suitable datasets from Twitter, filtered relevant tweets discussing vaccinations during the time frame from 01.01.2020 to 31.01.2022, and extracted relevant content. Specifically, we propose a semi-automatic analysis pipeline consisting of tweet filtering, sentiment analysis, and topic modeling to trace the rapid change of topics and public sentiment in the vaccination discourse. We further shed light on how the evolution of topics and sentiments relate to important policy events outlined in the national vaccination strategies of the three countries under investigation. Thus, we formulate the following research questions: \\
\begin{itemize}
    \item RQ1: How does the vaccination discourse in DACH countries on Twitter evolve (in terms of tweet frequencies and sentiments)? 
    \item RQ2: Which topics and themes were prevalent in the discourse? To what sentiments were they connected?
    \item RQ3: How did the topics, themes, and associated sentiments evolve?
    \item RQ4: How was this related to different phases of the pandemic and policy events?
\end{itemize}
Gaining insights into these research questions adds to the empirical literature on data-driven policy-making. Moreover, this study further advances the methodological literature on extracting, filtering, and analyzing Twitter data for policy research. 

We find that when the COVID-19 vaccines were first authorized, the debate on Twitter focused on a range of topics, including \emph{side-effects of individual vaccines} and \emph{vaccinations in general} but also \emph{freedom and civic liberties}. During later phases of the pandemic, when different policies restricting the freedom of unvaccinated citizens were publicly discussed and later implemented, the attention increasingly shifted away from medical and other concerns towards questions of \emph{freedom and civic liberties}. At the same time, vaccination uptake increased. 
This finding may indicate that these policies might have been an essential factor in attenuating vaccination hesitancy - either due to the imposed restrictions for the unvaccinated or/and the decreased attention to medical concerns. 
However, while vaccination hesitancy decreased, the discourse, which was connected to more negative than positive sentiments from the start, did not become more positive but, in fact, more polarized. 
This hints at the concerns being ignored by the citizens rather than having been resolved. 

\section{Related Work}
\subsection{Leveraging Discourse Data for Policy Making}
Discourse data can inform all stages of the policy cycles, from identifying the most pressing issues to evaluating policies after implementation. They have been successfully used to provide insights into citizens’ preferences for different policy alternatives \cite{CeronAndNegri2015, ceronAndNegri2016}. This information is vital to the formulation phase of public policy and contributes to more responsive policy-making. Regarding specific policy areas, citizens’ tweets were used in financial policy to predict consumers’ inflation expectations
\cite{AngelicoEtAl2022} and stock market indicators such as the Dow Jones, the NASDAQ, and the S$\&$P 500 \cite{bollenEtAl2011, ZhangEtAl2011}. Another use case for policy-relevant insights gained from SNS data pertains to emergency response in natural disasters. For instance, studies have shown that crowdsourcing people’s reactions on Twitter is a reliable, cheap, and scalable approach to detect extreme real-world events such as earthquakes \cite{vanQanEtAl2017, pobleteEtAl2018, sakakiEtAl2010}.

Insights gained from discourse data have also proven to be particularly fruitful with regard to health policies, for instance, to geographically locate illnesses such as allergies, obesity, and insomnia \cite{PaulAndDredze2021}. Furthermore, Twitter data can track and even forecast the spread of infectious diseases such as Influenza epidemics \cite{aramakiEtAl2011}
; Yang et al., 2021), and, more recently, COVID-19 \cite{kleinEtAl2021}. 
In addition, the Tweets analyzed in these studies could also be used to gauge public interest or concern about health-related events \cite{signoriniEtAl2011}, key information for policy-makers to introduce effective countermeasures.

\subsection{Twitter for Analyzing Public Opinion during the COVID-19 pandemic}

Several studies harvested data from Twitter to gain insights into public opinion about the pandemic.  \cite{jing2021characterizing} show that political actors in the US strategically used Twitter to establish partisan narratives about the pandemic. Moreover, verified Twitter users in Italy used their prominence to spread misinformation about the pandemic on the social media platform, especially those users associated with the right and center-right wing political community \cite{caldarelli2021flow}. Another study from the Chinese context suggests that public attention to the pandemic was largely influenced by specific events and measures introduced by the government \cite{cui2021attention}. 
\cite{wangEtAl2021} classified tweets as conforming with one of the four health belief models constructs using machine learning. They found that scientific events (e.g., publications) and nonscientific events (e.g., political speeches) seemed to have a comparable influence on health belief trends on Twitter. 

Several studies investigate prevalent topics and themes in the Twitter discourse about the COVID-19 pandemic \cite{xueEtAl2020, al-ramahiEtAl2021, Boon-IttAndYukolpat2020}. While \cite{xueEtAl2020} identified 13 topics in tweets through topic modeling and categorized them manually into five overarching themes (e.g., public health measures to slow the spread of the pandemic; social stigma associated with COVID-19), 
\cite{Boon-IttAndYukolpat2020} (the COVID-19 pandemic emergency, how to control COVID-19, reports on COVID-19) and \cite{al-ramahiEtAl2021, Boon-IttAndYukolpat2020} (constitutional rights and freedom of choice; conspiracy theory, population control, and big pharma; fake news, fake numbers, and fake pandemic) each find three overarching themes. All three studies developed lists of predefined hashtags to identify relevant tweets. While this procedure has its merits, it also risks missing other relevant tweets of the debate or inserting a selection bias. \cite{dooganEtAl2020} assigned 131 automatically generated topics to 22 non-pharmaceutical interventions grouped into seven categories: Personal Protection, Social Distancing, Testing and Tracing, Gathering Restrictions, Lockdown, Travel Restrictions, and Workplace Closures. They found that the proportion of intervention-related topics varied between the six investigated countries. The relationship between tweet frequencies and case numbers was statistically significant only for two countries. While less restrictive interventions gained widespread support, more restrictive ones were perceived differently across countries. 
\cite{RidhwanEtAl2021} used LDA to estimate a good number of topics to generate using Gibbs Sampling Dirichlet Multinomial Mixture, arriving at 35 topics. These were then manually labeled and assigned to 12 themes, such as lockdowns, social distancing, and travel and border restrictions. 
 
Many recent works are concerned with sentiment analysis or emotion detection in tweets about the pandemic. 
\cite{naseemEtAl2021} benchmark different sentiment analysis methods on their COVIDSENTI sentiment dataset, which consists of 90,000 COVID-19-related tweets covering February and March 2020, and find the transformer-based BERT to outperform all other methods. Their analysis shows that negative sentiments were prevalent and that intensity and frequency of negative sentiments were high before mid of March 2020 but decreased after. 

\cite{muellerEtAl2020} release COVID-Twitter-BERT, pre-trained on a large corpus of English COVID-19 tweets, which achieves considerable improvement over the BERT-LARGE base model on the target domain. 
\cite{shofiyaAndAbidi2021} use a hybrid approach based on SentiStrength and an SVM classifier for sentiment analysis and find that most users in Canada expressed neutral sentiment towards social distancing measures. 
\cite{xueEtAl2020} use scores based on a word-emotion association lexicon, and find as dominant emotions for their selected topics anticipation that measures can be taken, followed by mixed feelings of trust, anger, and fear related to different topics, especially when discussing new COVID-19 cases and deaths. 

Also using a lexicon-based approach, \cite{mathurEtAl2020} find that a high number of tweets is related to trust, which they interpret as people's confidence in the ability to fight COVID-19 and policies taken by authorities. At the same time, fear and sadness are also prevalent. They find the number of tweets with positive vs. negative emotions to be almost equal. 
\cite{RidhwanEtAl2021} use a Recurrent Neural Network for emotion classification and the lexicon-based tool VADER for sentiment analysis of COVID-19-related tweets relating to Singapore based on user information and geo-tags. 
Topics relating to measures such as social distancing and the encouragement to stay at home and to wear masks were coupled with positive sentiments. In contrast, negative sentiments dominated the discourse about travel and border restrictions. Overall, policies by the Singapore government were coupled with positive sentiments. The authors conclude that the citizenry  supported anti-COVID-19 measures.

\subsection{Twitter for Analyzing Public Opinion Towards COVID-19 vaccines}

Several studies zoom in more and specifically investigate public opinion about COVID-19 vaccines using Twitter data. \cite{huEtAl2021} structure the pandemic into phases according to pre-selected key events relating to the vaccination roll-out in the United States (US). The authors draw on geo-tagged tweets and find strong changes in public sentiment and emotion for the different phases. They conclude that social events and public announcements by influential entities may impact public opinion on COVID-19 vaccines considerably. \cite{FazelEtAl2021} investigate how positive vs. negative sentiments develop in relation to major news announcements about vaccines in the United Kingdom. They find that each announcement was associated with a short-term decrease in negative sentiment and that tweets with negative sentiment toward vaccines were posted by a smaller number of individuals. The high engagement created by negative tweets decreased gradually in the course of the vaccination campaign. Both studies start with the assumption that specific events drive change in public opinion and, thus, conducted a targeted search for events, such as public announcements and news coverage. 

The work by \cite{goranEtAl2021} analyzes reasons for vaccine hesitancy. Using a manually created list of keywords, the authors create a dataset of Twitter posts and accounts expressing a strong anti-vaccine stance. The findings indicate that vaccine hesitancy is fueled by misinformation originating from websites with questionable credibility. 
Furthermore, \cite{bonnevieEtAl2021} quantifies the rise of vaccine opposition on Twitter four months before and four months after the spread of the virus in the US. With a manually created keyword list, they collect tweets discussing vaccinations. The authors manually derive conversation themes of vaccination opposition from them, arriving at eleven themes, including adverse health impacts, policies and politics, and disease prevalence. They show that the frequency of these themes changed over time. 
\cite{sattarEtAl2021} perform sentiment analysis on tweets discussing vaccinations and propose a model to forecast vaccination uptake. The predicted numbers approximated the actual numbers for the US \cite{owidcoronavirus}. 
\cite{herrera-pecoEtAl2021} analyze a COVID-19 antivaccination campaign in Spanish tweets, one week before and one week after the European Medicines Agency announced the authorization of the Pfizer BioNTech vaccine. They find that attacks against vaccine safety were the most frequent antivaccine message. Moreover, the authors also find conspiracy theories, such as presenting the vaccine as a means of manipulating the human genetic code. 

Our studies draw on the insights these studies put forth and apply them to the German-speaking context. Moreover, we deviate from these studies by introducing data-driven approaches to identify relevant tweets, peaks, and change points, inserting as little prior knowledge and assumptions as possible into the analyses.  

\section{Methods}

\subsection{Dataset and Preprocessing} 
We examine tweets within the time span of 01.01.2020 - 31.01.2022 using the \emph{TweetsKB} \cite{fafaliosEtAl2018} pipeline. TweetsKB is a large-scale knowledge base of annotated tweets harvested using the Twitter streaming API. Since 2013, a random 1\% sample of the Twitter stream has been harvested. Their metadata and information automatically extracted from the tweets, such as entities, sentiments, hashtags, and user mentions, is accessible in RDF format. The data from 2013 until December 2020 is available at \url{https://data.gesis.org/tweetskb/}. We use the same pipeline to harvest tweets for the above-mentioned period, resulting in 12,297,163 tweets. 
For our analyses, we analyze the textual content of tweets, ignoring pictures and links to videos or other content. 

\subsubsection{Relevance Filtering} 
We extract relevant tweets by filtering for time ("01.01.2020 - 31.01.2022"), language ("German"), and topic ("vaccinations"). 
We use the timestamps provided by Twitter to filter tweets created during our desired time frame. Furthermore, we draw on Twitter's language tag to filter German tweets. 
Ensuring that the tweets address the relevant topic of vaccinations is not trivial. It requires the complex procedure of creating a seed list with relevant search strings. Commonly, researchers rely on manually created seed lists for hashtags or search strings to identify relevant tweets, e.g., \cite{RidhwanEtAl2021, buntain2018sampling, goranEtAl2021, bonnevieEtAl2021, xueEtAl2020, al-ramahiEtAl2021, herrera-pecoEtAl2021}.  
However, manual seed list creation is costly and lowers the reproducibility of results as seed lists for similar topics exhibit a high variance with unknown effects on generated results. For example, the keyword list used to filter COVID-19 - related tweets in \cite{chenEtAl2020} comprises 80\footnote{the list is updated by the authors: \url{https://github.com/echen102/COVID-19-TweetIDs/blob/master/keywords.txt}, last update to date was on 11/28/2021} keywords. \cite{dimitrovEtAl2020} expanded this list to include 268 keywords for their TweetsCov-19 dataset, while the multilingual keyword list used by \cite{imranEtAl2022} for the TBCOV corpus includes more than 800 terms. 
Moreover, manually curated lists may inadequately capture vocabulary mismatch problems, emerging new terms and are prone to biases - e.g., focusing on certain topics or frames while neglecting others. 

Since we are interested in exploring different topics assuming as little prior knowledge as possible, we followed an automatic query term expansion approach to generate a list of search terms (\emph{seed list}).
Starting with an initial query keyword (“Impfung”, English: “vaccination”), we extracted all tweets that contain this keyword as a single token word while not applying case sensitivity. We created a set of candidate terms from this set of tweets by collecting and lemmatizing all verbs, adjectives, nouns, and proper nouns using the Spacy POS tagger \cite{spacy}.

Next, we reduced the set of candidates by removing all terms that fall under a given limit of semantic similarity compared to the query keyword. To determine said similarity, we used pre-trained word embeddings from Fasttext.cc trained on Wikipedia and Common Crawl \cite{fasttextmethod}; precisely, we used the German dataset with 300 dimensions \cite{fasttextembeddings}. On this embedding, we computed the cosine-similarities between the query keyword ("Impfung") and the candidate keywords. The similarities range between $-1$ and $1$. Visual inspection of the candidate terms indicates that a minimum cosine-similarity of $0.6$ is required to retrieve meaningful results. 

The remaining candidate terms were then sorted by the number of their co-occurrence with the query keyword, and we selected the top 30 terms for the seed list. 

Table \ref{tab:terms} displays the resulting sets of seed terms. 
\begin{table}[]
    \centering
    \begin{tabular}{l|l}
        \textbf{Seed Term} & \textbf{Translation} \\
        impfung & vaccination \\
        impfen & to vaccinate \\
        impfstoff & vaccine \\
        geimpfte & (the) vaccinated \\
        impfungen & vaccinations \\
        infektion & infection \\
        impfpflicht & compulsory vaccination \\
        geimpft & vaccinated \\
        impfschutz & immunization protection \\
        impftermin & vaccination appointment \\
        impfschäden & vaccination damages \\
        impfschaden & vaccination damage \\
        immunisierung & immunization \\
        impfkampagne & vaccination campaign\\
        masern & measles \\
        impfnebenwirkungen & vaccination side-effects \\
        impfling & freshly vaccinated person / seed chrystal \\
        erstimpfung & primary vaccination \\
        grippeimpfung & influenza vaccination \\
        impftermine & vaccination appointments \\
        impfreaktionen & reactogenicities \\
        impfreaktion & reactogenicity \\
        impf & vaccination- \\
        auffrischungsimpfung & booster injection \\
        zwangsimpfung & compulsory vaccination \\
        impfaktion & vaccination event \\
        schweinegrippe & swine flu\\
        impfstoffs & vaccine (direct object)\\
        grundimmunisierung & fundamental immunization \\
        impfbereitschaft & vaccination willingness \\
    \end{tabular}
    \caption{Automatically generated seed list for filtering tweets discussing vaccinations and added English translations }
    \label{tab:terms}
\end{table}

This automated procedure suggested including terms referring to other viruses than Corona, e.g., the swine flu. As we assumed such discourses to relate to discourse about COVID-19 in the selected time frame, we do not exclude these keywords from our list. 

To construct our final set of tweets discussing vaccinations, we searched for all keywords in the set in all tweets harvested by our TweetsKB pipeline in the specified time frame. We then extracted all tweets mentioning at least one of the keywords in their tweet texts or hashtags. This resulted in a set of 201,705 tweets. Removing all tweets written in a language other than German according to the language tag provided by Twitter, resulted in our final set of 199,207 tweets. Revisiting the list of keywords, we expect the seed term "Infektion" (engl. infection) to add noise to our data because such tweets do not necessarily address vaccinations. 
We excluded 7457 tweets (3.89\%) due to the occurrence of this term (alone) from our analysis.

\subsubsection{Sentiment Analysis}
We use the automatic tool SentiStrength to identify tweet sentiments. It is tailored for the analysis of short social media texts \cite{ThelwalEtAl2012} and  measures the strength of both positive and negative sentiments in a tweet on a scale from 1 to 5. Thus, every tweet has one score specifying the intensity of the negative and one score specifying the intensity of the positive sentiment. 

Based on the automatically assigned sentiment scores and the tweets’ timestamps, we generate time series data, accumulating all sentiments for one day using four different approaches: 1) summing up all sentiment scores (positive and negative intensity scores) per day (SUM), 2) normalizing the summed up score by the number of tweets (REL) and 3) counting the number of positive (POS) and 4) negative (NEG) tweets for each day. A tweet is considered positive when the intensity of its positive sentiment is higher than the intensity of its negative sentiment and vice versa. 
Note that for generating the plots, we translate the sentiment scores to intervals of 0 to 4 and -4 to 0, respectively. Using the SUM and REL metrics, intensities of sentiments are being regarded, while for POS and NEG, sentiment intensities are translated to positive, negative, and neutral/mixed labels without any information on intensity. All metrics except REL represent the frequency of tweets in addition to the sentiments. 
Note that by summing up intensity scores, we do not differentiate between tweets that have a neutral sentiment, i.e. no negative and no positive sentiment, and tweets that have a mixed sentiment with both negative and positive sentiments being equally strong. 

\subsection{Topic Modeling} \label{sec:topics} 
We use BERTopic \cite{Grootendorst2022}, a recent transformer-based topic modeling technique, to derive topics from the tweet texts in an unsupervised manner, i.e., all topics are derived from the data without relying on any prior knowledge. We decided to use embedding-based topic modeling instead of “traditional” topic modeling techniques such as LDA \cite{Blei.2003.lda}. This approach enables us to exploit information about semantic relationships among words, represented in embeddings generated on large data volumes instead of relying on the distribution of words in our tweets alone. 
Traditional techniques cluster co-occurring words to find topics and then proceed to identify these topics in a set of documents. Input is typically pre-processed (e.g., using lemmatization), and information about sentence structure is disregarded in favor of treating documents as bags of words. Embedding-based methods such as BERTopic, in contrast, typically do not preprocess or otherwise alter the input. They consider the semantic similarity of documents using embeddings to cluster similar documents into topics and try to find typical terms that characterize them in a separate step. 
BERTopic allows using custom embeddings. We use embeddings-paraphrase-multilingual-MiniLM-L12-v2
sentence transformers model \footnote{\url{https://huggingface.co/sentence-transformers/paraphrase-multilingual-MiniLM-L12-v2}}, a multilingual sentence-transformers model which maps sentences and paragraphs to a 768-dimensional dense vector space. It was trained on parallel data for 50+ languages and proved useful for semantic search and clustering.
Using these embeddings allows us to find similarities in sentences within one language or across languages, a valuable property for German tweets that may use English terms or quote English content. 
We use BERTopic's default algorithms UMAP \cite{umap} to reduce the dimensionality of the document embeddings, and HDBSCAN \cite{hdbscan} for document clustering. 

We compute topics for the complete set of tweets and then classify them into negative and positive tweets to gain insights into their occurrence in different contexts. 
Each tweet is assigned to precisely one topic, with one noisy residual category for all tweets that do not fit into any of the topic clusters with high probability. 
Note that, in principle, one tweet may be assigned to more than one topic based on the calculated probabilities of a tweet belonging to any cluster. 
Due to the limited length of tweets, we keep the standard procedure of assigning only the most prevalent topic. 

Since we use the 1\% Twitter API, one tweet in our corpus mentioning a topic represents a much larger discourse. We thus decided to keep the standard value of ten documents for the minimum topic size and set the number of topics to 150 to enable a fine-grained analysis while maintaining a number of topics that is feasible to review manually. By grouping topics into themes, we perform a manual merging step later on. Therefore we prefer a high number of topics at this step to prevent information loss. 

For optimizing topic representation after extraction, we set the ngram range to 1,2 and the diversity to 1.0. However, we did not rely on the extracted topic representations alone when interpreting the clusters. Instead, we manually assigned labels to each topic by examining the tweets in the respective clusters. 
For this, the first two authors of this paper (one Computer Scientist and one Political Scientist) labeled all clusters independently and discussed their results. 
For some of the topics, the labels diverged regarding their precise wording, but not regarding the perceived content. 
Final labels were assigned by both authors jointly. 
For 11 topics we failed to find suitable labels as the tweets seemed too heterogeneous. We excluded these clusters from our analysis. 

\subsection{Phases of the pandemic and policy events}
\begin{table}[ht!]
    \centering
    \begin{tabular}{l|l|l|l|l|l|l}
        \textbf{Phase}&\textbf{Begin}&\textbf{End}&\textbf{\% D}&\textbf{\% A}&\textbf{\% CH}&\textbf{\% UK}\\
         Sporadic cases&27.01.2020&24.02.2020&&&&\\
         Wave 1&02.03.2020&18.05.2020&&&&\\
         Summer plateau 2020&18.05.2020&28.09.2020&&&&\\
         Wave 2&28.09.2020&01.03.2021&5.2&5.2&6.4&30\\
         Wave 3&01.03.2021&14.06.2021&49&48&44&62\\
         Summer plateau 2021&14.06.2021&02.08.2021&63&60&55&70\\
         Wave 4&02.08.2021&27.12.2021&75&74&69&77\\
         Wave 5&27.12.2021&31.01.2022*&77&76&70&78 \\
    \end{tabular}
    \caption{Phases of the pandemic as classified by the RKI (own translation of the phase labels; calendar weeks mapped to dates) with added vaccination ratio statistics provided by Our World in Data. \%D/A/CH/UK: percentage of vaccinated people in Germany/Austria/Switzerland/UK, respectively, at the end of the phase  *end of the investigated time frame}
    \label{tab:rkiPhases}
\end{table}

To relate the evolution of the discourse to different phases of the pandemic, we refer to the classification provided by the Robert Koch Institute (RKI) \footnote{\url{https://www.rki.de/DE/Content/Infekt/EpidBull/Archiv/2022/Ausgaben/10_22.pdf?__blob=publicationFile}}, the German government’s central scientific biomedicine institution. The phases from the beginning of the pandemic until the end of the time under investigation in this study are classified as listed in Table \ref{tab:rkiPhases}. To compare the vaccination uptake in Germany, Austria, and Switzerland for each of the different phases, we add vaccination ratios provided by Our World in Data \cite{owidcoronavirus}. Even though the RKI classification refers to the spread of the virus in Germany, \cite{DessonEtAl2020} show that the German-speaking countries faced similar epidemiological situations during the pandemic. 
We draw on official websites (e.g., Bundestag.de, zusammengegencorona.de, sozialministerium.at) and Wikipedia to identify events relating to vaccination policies in the DACH countries, such as the licensing of new vaccines. We arrive at a list of 57 events: 22 for Germany (Table \ref{tab:eventsD}), 16 for Switzerland (Table \ref{tab:eventsCH}), and 19 for Austria (Table \ref{tab:eventsA}). 
The tables show that the three countries partly issued similar policies at similar times. 
These do not fully coincide with the pandemic phases. 
To analyze the discourse about policy events, we derive \textit{policy phases} by grouping similar policy events (see Table \ref{tab:policyPhases}).

\begin{table}[ht!]
    \centering
    \begin{tabular}{p{1.5cm}|p{5.6cm}|p{5cm}|p{1.0cm}|p{0.8cm}}
        \textbf{Date}&\textbf{Event}&\textbf{Source}&\textbf{RKI Phase}&\textbf{Policy Phase}\\
         09/11/2020&Leopoldina publishes position paper on vaccine distribution&\url{https://www.leopoldina.org/}&Wave 2&I\\
         17/12/2020&First publication of the vaccination recommendation (STIKO)&\url{https://edoc.rki.de/handle/176904/7579}&Wave 2&II\\
         18/12/2020&Vaccination sequence published in the Federal Gazette&\url{bundesanzeiger.de}&Wave 2&II\\
         21/12/2020&Authorization of BioNTech vaccine&\url{https://impfdashboard.de/}&Wave 2&II\\
         27/12/2020&Vaccination start&\url{https://impfdashboard.de/}&Wave 2&II\\
         06/01/2021&Authorization of Moderna vaccine&\url{https://impfdashboard.de/}&Wave 2&II\\
         29/01/2021&Authorization of AstraZeneca vaccine&\url{https://impfdashboard.de/}&Wave 2&II\\
         11/03/2021&Authorization of Johnson \& Johnson vaccine&\url{https://impfdashboard.de/}&Wave 3&II\\
         15/03/2021&Halt of AstraZeneca vaccinations due to safety concerns&
         \url{https://www.bundesregierung.de}&Wave 3&II\\
         25/03/2021&Resumption of AstraZeneca vaccinations&
         \url{https://www.bundesregierung.de}&Wave 3&II\\
         07/04/2021&Nationwide vaccination in doctors' offices&\url{https://impfdashboard.de/}&Wave 3&II\\
         06/05/2021&Nationwide suspension of priority groups for AstraZeneca&\url{https://www.ndr.de/}&Wave 3&III\\
         28/05/2021&Authorization of BioNTech vaccine for youth&\url{https://investors.biontech.de/}&Wave 3&IV\\
         07/06/2021&Official end of vaccination priority groups&\url{https://www.tagesschau.de}&Wave 3&IV\\
         23/08/2021&Implementation of 3G Rule restricting access to facilities for unvaccinated and untested individuals&\url{https://www.bundesregierung.de}&Wave 4&IV\\
         05/11/2021&Health ministry decides on offering booster shots&\url{https://www.bundesregierung.de/}&Wave 4&V\\
         26/11/2021&Authorization of BioNTech vaccine for children&\url{https://www.pei.de/}&Wave 4&V\\
         01/12/2021&Vaccination stop for AstraZeneca vaccine&\url{https://www.zusammengegencorona.de/}&Wave 4&V\\
         10/12/2021&Adoption of facility-based mandatory vaccination&\url{https://www.bgbl.de/}&Wave 4&V\\
         21/12/2021&STIKO recommends shorter time span before booster shot&\url{https://www.rki.de/}&Wave 4&V\\
         21/12/2021&EU Commission decides on limited validity of vaccination certificates&\url{http://data.europa.eu/}&Wave 4&V\\
         22/02/2022&Authorization of Novavax vaccine&\url{https://www.pei.de/}&Wave 5&V\\
    \end{tabular}
    \caption{List of policy events: Germany}
    \label{tab:eventsD}
\end{table}

\begin{table}[ht!]
    \centering
    \begin{tabular}{p{1.5cm}|p{5.6cm}|p{5cm}|p{1cm}|p{0.8cm}}
        \textbf{Date}&\textbf{Event}&\textbf{Source}&\textbf{RKI Phase}&\textbf{Policy Phase}\\
         25/11/2020&Chancellor's office statement on ethical issues of vaccination published&\url{https://www.bundeskanzleramt.gv.at/}&Wave 2&I\\
         21/12/2020&Authorization of BioNTech vaccine&\url{https://www.basg.gv.at/}&Wave 2&II\\
         26/12/2020&Publication of priority groups&\url{https://web.archive.org/}&Wave 2&II\\
         27/12/2020&Start of vaccinations&\url{https://www.wienerzeitung.at/}&Wave 2&II\\
         06/01/2021&Authorization of Moderna vaccine&\url{https://www.basg.gv.at/}&Wave 2&II\\
         28/01/2021&Regulation on the procedure of Corona vaccination enters into effect&\url{https://www.ots.at/}&Wave 2&II\\
         29/01/2021&Authorization of AstraZeneca vaccine&\url{https://www.basg.gv.at/}&Wave 2&II\\
         01/02/2021&Vaccination plan published&\url{https://www.sozialministerium.at/}&Wave 2&II\\
         11/03/2021&Authorization of Johnson \&d Johnson vaccine&\url{https://www.basg.gv.at/}&Wave 3&II\\
         25/05/2021&BioNTech recommended for persons older than 11&\url{https://www.basg.gv.at/}&Wave 3&IV\\
         23/07/2021&Moderna recommended for persons older than 11&\url{https://www.basg.gv.at/}&Summer plateau 2021&IV\\
         08/10/2021&BioNTech booster recommended for from 6 months after second jab&\url{https://www.basg.gv.at/}&Wave 4&IV\\
         23/10/2021&Introduction of lockdown for the unvaccinated upon reaching 600 patients in intensive care&\url{https://www.wienerzeitung.at/}&Wave 4&IV\\
         29/10/2021&Moderna booster recommended for from 6 months after second jab&\url{https://www.basg.gv.at/}&Wave 4&IV\\
         15/11/2021&Nationwide lockdown for the unvaccinated&\url{https://www.wienerzeitung.at/}&Wave 4&V\\
         25/11/2021&Authorization of BioNTech vaccine for children aged 5 - 11&\url{https://www.basg.gv.at/}&Wave 4&V\\
         17/12/2021&Johnson \& Johnson booster recommended for from 6 months after second jab&\url{https://www.basg.gv.at/}&Wave 4&V\\
         20/12/2021&Authorization of Novavax vaccine&\url{https://www.basg.gv.at/}&Wave 4&V\\
         20/1/2022&Introduction of compulsory vaccinations for all citizens older than 17&\url{https://www.wienerzeitung.at/}&Wave 5&V\\
    \end{tabular}
    \caption{List of policy events: Austria}
    \label{tab:eventsA}
\end{table}

\begin{table}[ht!]
    \centering
    \begin{tabular}{p{1.5cm}|p{5.6cm}|p{5cm}|p{1cm}|p{0.8cm}}
        \textbf{Date}&\textbf{Event}&\textbf{Source}&\textbf{RKI Phase}&\textbf{Policy Phase}\\
         17/12/2020&Vaccination sequence published by the BAG&\url{https://www.bag.admin.ch/}&Wave 2&II\\
         19/12/2020&Authorization of BioNTech vaccine&\url{https://www.admin.ch/}&Wave 2&II\\
         23/12/2020&Start of vaccinations&\url{https://www.srf.ch/}&Wave 2&II\\
         12/01/2021&Authorization of Moderna vaccine&\url{https://www.swissmedic.ch/}&Wave 2&II\\
         13/01/2021&Vaccination costs covered by statutory health insurance approved&\url{https://www.bag.admin.ch/}&Wave 2&II\\
         21/01/2021&Zurich is one of the first regions to start involving primary care doctors&\url{https://telebasel.ch/}&Wave 2&II\\
         03/02/2021&Swissmedic requests further data for approval for AstraZeneca&\url{https://www.swissmedic.ch/}&Wave 2&II\\
         22/03/2021&Authorization of Johnson \& Johnson vaccine&\url{https://www.swissmedic.ch/}&Wave 3&II\\
         22/04/2021&Work on international vaccination certificate begins&\url{https://www.bag.admin.ch/}&Wave 3&III\\
         01/06/2021&Those who have recovered should also be vaccinated&\url{srf.ch/}&Wave 3&IV\\
         04/06/2021&Legal basis for issuance of vaccination certificates created&\url{https://www.bag.admin.ch/}&Wave 3&IV\\
         26/10/2021&Booster vaccination recommended for persons older than 65&\url{https://www.bag.admin.ch/}&Wave 4&IV\\
         04/11/2021&Authorization of AstraZeneca not further pursued&\url{https://www.swissmedic.ch/}&Wave 4&V\\
         26/11/2021&Booster vaccination recommended for the general population&\url{https://www.bag.admin.ch/}&Wave 4&V\\
         10/12/2021&Authorization of BioNTech vaccine for children&\url{https://www.bag.admin.ch/}&Wave 4&V\\
         21/12/2021&Recommendation to shorten the time until the booster vaccination&\url{https://www.bag.admin.ch/}&Wave 4&V\\
    \end{tabular}
    \caption{List of policy events: Switzerland}
    \label{tab:eventsCH}
\end{table}

\begin{table}[ht!]
    \centering
    \begin{tabular}{p{1cm}|l|l|p{7cm}}
        \textbf{Policy Phase}&\textbf{Begin}&\textbf{End}&\textbf{Description}\\
         I&01/11/2020&10/12/2020&Beginning of the official COVID-19 vaccination policies\\
         II&10/12/2020&15/04/2021&Publishing of vaccination strategies, authorization of the vaccines and vaccination start, halt, and resumption of AstraZeneca vaccinations in Germany\\
         III&15/04/2021&15/05/2021&Suspension of priority groups for AstraZeneca vaccines in D; international vaccination certificate preparations in CH\\
         IV&15/05/2021&01/11/2021&Vaccine recommendations for specific age groups, access restrictions for unvaccinated persons in Germany\\
         V&01/11/2021&30/01/2021&Booster shot recommendations and authorizations of vaccines for children; AstraZeneca vaccination stop in Germany; lockdowns for unvaccinated under certain conditions in Austria\\
    \end{tabular}
    \caption{Policy phases for the DACH countries}
    \label{tab:policyPhases}
\end{table}

\subsection{Detection of Peaks in Tweet Frequencies} \label{sec:peaks} 
We define a peak as a point in time where the respective value deviates from the expected interval (mean +/- standard deviation (std)) by more than 1.5 times the expected maximum or minimum value. 
\begin{equation*}
    ((mean + std) + (|(mean + std)| *1.5) > peak < ((mean - std) - (|(mean - std)| *1.5|)
\end{equation*}

\subsection{Detection of Change Points} \label{sec:changePoints} 
For detecting points in which the tweet frequency changes in the time series data, arguably due to shifts in public attention or opinion, we use the Python library ruptures \cite{TruongEtAl2020}. It includes several offline change detection methods for non-stationary signals. We opt for the Pelt (Penalized change point detection) search algorithm, which does not require setting a fixed number of change points in advance. This implementation computes the segmentation, which minimizes the constrained sum of approximation errors for a given model and penalty level \cite{KillickEtAl2012}. We use Pelt with the ruptures standard parameters.

\subsection{Detection of Trends} \label{sec:trends} 
We employ the Mann-Kendall test \cite{Mann1945, Kendall1975} to determine whether significant trends in the data regarding sentiments and tweet frequencies exist. We use the Original Mann-Kendall test supplied by \cite{Hussain2019pyMannKendall}\footnote{\url{https://pypi.org/project/pymannkendall/}}. This non-parametric test does not consider serial correlation or seasonal effects. The standard alpha significance level is set at 0.05.

\section{Data Analysis and Results}

\subsection{Evolution of Vaccination Discourse in DACH Countries} 

To investigate RQ1, we analyze the development of tweet frequencies and sentiments in the vaccination discourse over time and relate them to the general German Twitter discourse. 

\begin{figure}[ht]
    \includegraphics[width=\textwidth]{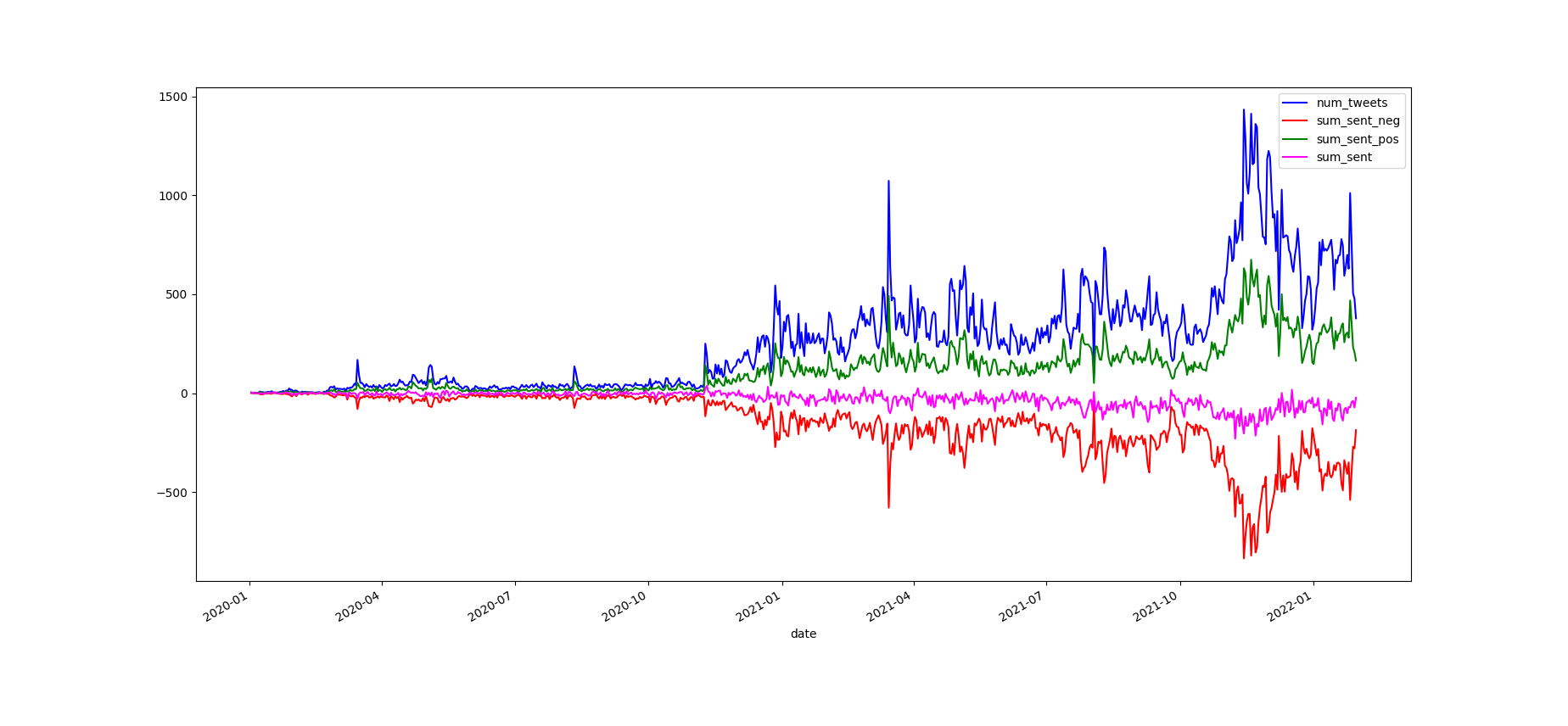}
    \caption{Frequencies and summed up sentiments of vaccination tweets over time. The blue line indicates the number of tweets on a given day, the green line the summed-up positive, the red line the summed-up negative, and the magenta line the overall summed-up sentiment intensities, respectively. }
    \label{fig:summedUpSentimentsOfTweets}
\end{figure}

\begin{figure}[ht]
    \centering
    \includegraphics[width=\textwidth]{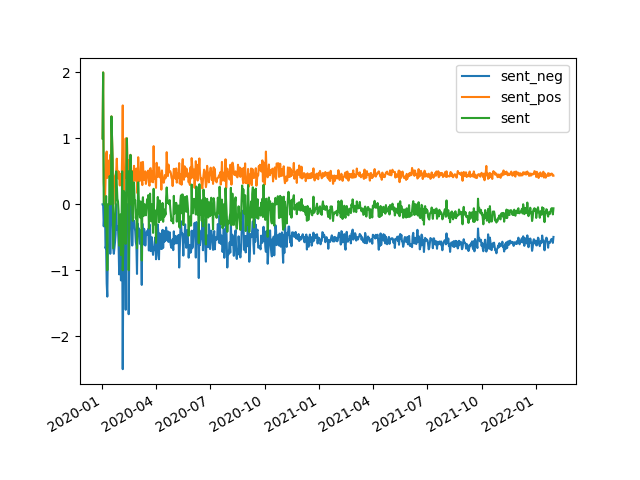}
    \caption{Relative sentiment of vaccination tweets over time. Orange marks the relative positive sentiment, blue the relative negative and orange the relative summed up positive and negative sentiment intensities, respectively. }
    \label{fig:relSentimentOfTweets}
\end{figure}

As Figure \ref{fig:summedUpSentimentsOfTweets} illustrates, before December 2020, only very few tweets mention any vaccination-related terms. This affirms that the vaccination discourse captured by our automatically generated seed list is indeed driven by COVID-19 vaccinations. Figure \ref{fig:relSentimentOfTweets} plots the REL sentiment scores, i.e., they are normalized with regard to the number of tweets. Both figures reveal that the overall vaccination discourse shows stronger negative than positive sentiments. Moreover, the discourse becomes slightly more negative over time. Thus, the negative sentiments were more negative than the positive sentiments were positive. Both the plotted sentiment and tweet frequencies hint at strong fluctuations over time. 

Trend analysis using the Mann-Kendall test reveals a significant decreasing trend both for the relative overall sentiment and the relative negative sentiment intensities. We find no trend for the relative positive sentiment intensities. This means that the negative sentiments became more negative over time while the intensity of positive sentiments remained constant.  

However, while the summed-up sentiment intensities (SUM) are more negative than positive, the number of predominantly positive tweets is slightly higher than the number of predominantly negative tweets: 53,176 positive tweets (26.69\%) and 49,342 (24.77\%) negative tweets of 199,207 tweets in total (including tweets with neutral/mixed sentiment). Thus, negative sentiments seem to be expressed with higher intensity than positive tweets. Yet, with a mean of -0.09, the average relative sentiment is close to neutral/mixed. 
The number of positive and negative tweets as well as the overall tweet frequency increase significantly with time. 

\begin{figure}[ht]
    \centering
    \includegraphics[width=\textwidth]{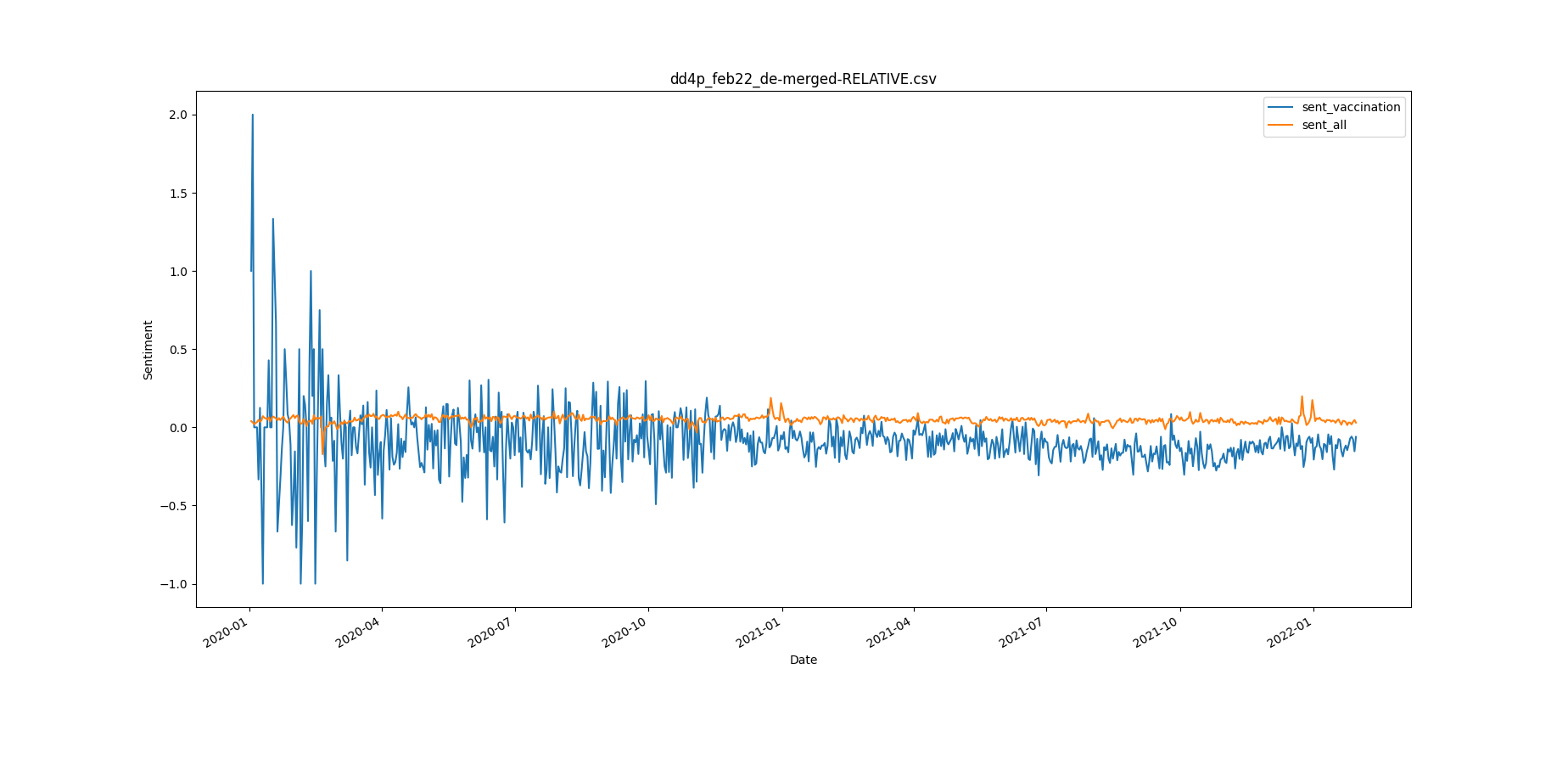}
    \caption{Sentiment in German tweets (orange) vs. sentiment in German vaccination tweets (blue).}
    \label{fig:sentiments}
\end{figure}

We further investigate whether the negative sentiments are inherent to the vaccination discourse or due to an overall negative German Twitter discourse. For that, we analyze the sentiments for all German tweets harvested with our pipeline during the investigated time frame. 
Figure \ref{fig:sentiments} relates the sentiments in the German vaccination tweets to the sentiments in German tweets of all topics in the same time frame. Depicted sentiments are the REL scores. The strong fluctuations in sentiment at the beginning of the year 2020 for the vaccination-related tweets can be attributed to the relatively low number of tweets in that time frame (see Figure \ref{fig:summedUpSentimentsOfTweets}). 
Similarly, the vaccination sentiments seem to exhibit higher fluctuations due to the smaller number of tweets compared to the general Twitter discourse. 
The results show that the discourse about vaccinations is more negative than the general discourse in German tweets. 
In the latter, the sentiment was overall more positive than negative, both in terms of summed-up and relative sentiment intensities, with a mean of 0.05 for the REL score (as compared to -0.09 for the vaccination tweets) and in terms of numbers of tweets with 1,758,776 (14.30\%) negative and 3,015,915 (24.53\%) positive of a total of 12,297,163 tweets. 
There is also a significant negative trend for the general German Twitter discourse, which is caused by both the negative sentiments becoming more negative and the positive sentiments becoming less positive, according to the Mann-Kendall trend analysis. 
Also for the general discourse, numbers for negative, positive, and all tweets show a significant increasing trend.

While significant trends can be observed over time, both the tweet frequencies and sentiments also fluctuate heavily at different points in time. 
Also, while the summed-up positive and negative sentiments are relatively close to balancing each other out, the increasing trend for positive sentiments and the decreasing trend for negative sentiment scores suggest that the discourse is indeed rather emotional and increasingly so.  

To get insights into what happened at different points in time, the following sections investigate in more detail which topics were discussed, in general, and when tweet frequencies increase, i.e. which topics are in the focus of attention, and which topics were responsible for positive and negative sentiments and sentiment trends. 

\subsection{Topics, Sentiments, and Themes} 
To answer RQ2, 
we first investigate the topics generated for all tweets as described in the \textit{Topic Modeling} Section and rank them by their frequencies. 

\subsubsection{Topics and Associated Sentiments}

The top 30 topics are depicted in Table \ref{tab:topicChart}. 

\begin{table}[ht!]
    \begin{tabularx}{\textwidth}{l|X|c|c}
      \textbf{Rank}&\textbf{Topic Label}&\textbf{$\#$}&\textbf{\ensuremath\heartsuit}\\
      \hline
    1&Children&8892&-0.12\\\hline
    2&Anecdotes: Experience with Corona vaccination&2210&-0.17\\\hline
3&(Prominent) vaccinated and unvaccinated men&2058&-0.07\\\hline
4&Situation in Germany and comparisons&1796&-0.07\\\hline
5&'Do (not) get vaccinated'&1738&0.16\\\hline
6&Corona and other flu viruses&1600&-0.25\\\hline
7&Corona in Israel&1533&-0.20\\\hline
8&Regulations&1476&-0.04\\\hline
9&AstraZeneca vaccine&1425&-0.01\\\hline
10&Duration of vaccination protection&1382&0.04\\\hline
11&Lockdowns&1132&-0.10\\\hline
12&Mutations and virus spread when vaccinated&1118&-0.11\\\hline
13&Basic rights&1093&0.14\\\hline
14&Practical implementation&1055&0.39\\\hline
15&Propaganda and fake news&1021&-0.34\\\hline
16&Vaccinations for elderly people&988&-0.13\\\hline
17&Compulsory vaccination&938&-0.15\\\hline
18&Metadiscussion about Twitter vaccination discourse&761&-0.15\\\hline
19&Masks and mask mandate&729&-0.15\\\hline
20&Vaccine efficacy for Omicron&676&-0.17\\\hline
21&Compulsory vaccination at work&651&-0.06\\\hline
22&SARSCoV2&635&-0.12\\\hline
23&Demonstrations and protests&633&-0.04\\\hline
24&Merkel&625&-0.07\\\hline
25&Vaccinations for children: medical views&611&-0.08\\\hline
26&mRNA vaccines&567&-0.02\\\hline
27&Immune system&550&-0.19\\\hline
28&Statistics about vaccination uptake and policies&535&-0.08\\\hline
29&\#AllesInDenArm&476&-0.01\\\hline
30&Austria&472&0.00\\\hline
    \end{tabularx}
    \caption{Top 30  topics (complete time interval. $\#$ Number of tweets \ensuremath\heartsuit Average sentiment per tweet (REL)}
    \label{tab:topicChart}
\end{table}

The vaccination discourse covers a wide range of topics (Table \ref{tab:topicChart}). 
The most frequently discussed topic addresses the question of whether  \textit{children} should be vaccinated, how they can be protected, their role in transmitting the virus, and their role in the pandemic more generally. A high number of tweets share personal experiences with Corona vaccinations (\textit{Anecdotes: Experience with Corona vaccination}), be it the authors' own or those of their peers, and discussing (prominent) (un)vaccinated persons, mostly men (\textit{(prominent) vaccinated and unvaccinated men}). 
Analyzing the topic clusters, we find that the topic modeling algorithm encoded the gender of the individuals. The first topic featuring anecdotal information focuses on females, while this one mainly contains tweets about males, many prominent individuals among them. 
Specific influential individuals are discussed in other topics. The highest rankings are Germany's chancellor at the time, Angela \textit{Merkel}, and Germany's current health minister Karl \textit{Lauterbach}, ranking 24th and 31st, respectively. 
Since the beginning of the pandemic, Lauterbach frequently appeared in German TV shows and interviews, voicing his opinion about measures, policies, and possible developments. 
A country-level view of the pandemic ranks fourth (\textit{Situation in Germany and comparisons}), focusing on the situation in Germany and comparing it with other countries. Other topics focusing on the country level are those debating Corona and the measures taken in Israel (\textit{Corona in Israel}, rank 7) and \textit{Austria} (rank 30). 
Twitter has also been used frequently for mobilizing others: a topic cluster containing calls to get vaccinated or not (\textit{'Do (not) get vaccinated'}) ranks fifth, the \textit{\#AllesInDenArm} ("everythingIntoTheArm") campaign 29th. 
Prominent individuals and ordinary Twitter users used this hashtag to communicate their vaccination status and motivate others to get vaccinated. 
Comparing COVID-19 infections with the flu and discussing experiences with the swine flu and flu vaccinations (\textit{Corona and other flu viruses}) also received significant attention, ranking this topic sixth. 
This topic also includes tweets discussing the severity of COVID-19 infections and whether the classification as a pandemic is justified. 

Rules and \textit{Regulations} are also widely discussed. This topic contains tweets about the so-called 2G and 3G rules which restricted access to certain facilities to vaccinated, recovered, or negatively tested individuals (rank 8). 
Other regulations that were frequently debated were \textit{Lockdowns} (rank 11), \textit{Masks and mask mandates} (19). 
Discourses about \textit{Compulsory vaccinations} reached rank 17, and about \textit{Compulsory vaccinations at work} specifically rank 21. 
Relatedly but assigned to a separate topic, we find a more general debate about freedom, personal responsibilities, restrictions, and \textit{Basic rights} (13). Closely related is the topic of \textit{Demonstrations} and protests (23). 
The vaccine that seemed to spark the most discussions is the \textit{AstraZeneca vaccine}, ranking 9th among all topics. 
\textit{mRNA vaccines}, i.e., gene-based vaccines, belong to the topic cluster ranking 26th, which evolves primarily around their mode of operation. 
Other vaccines are clustered into separate topics, too, but do not rank among the 30 most frequent topics (Sputnik: 42, BioNTech: 56, Novavax: 81, Johnson \& Johnson: 100 and, when referred to as 'J\&J' 148, Moderna: 101). 
Tweets dealing with the efficacy of vaccinations can be found in many topics in the top 30. The \textit{Duration of vaccination protection} ranks 10th. \textit{Mutations and virus spread when vaccinated} at rank 12 refers to the extent to which the virus can be spread by the vaccinated and whether vaccinations pander to the evolution of mutations. \textit{Vaccine efficacy for Omicron} is discussed at rank 20. 
A more general debate about the influence of vaccinations and the Coronavirus on the \textit{Immune system} ranks 27th. This topic also includes debates about immunization through a COVID-19 infection compared to immunization through vaccination. 
\textit{Vaccinations for elderly people} are discussed in the topic ranking 16th, \textit{Vaccinations for children: medical views} at rank 25. Note that the latter topic focuses on medical considerations while other aspects can be found in the more general \textit{Children} topic ranking first. 
The \textit{Practical implementation} of the vaccination rollout also receives much attention on Twitter (rank 14). This topic includes tweets regarding opportunities for getting vaccinated.

Meta discussions about the news coverage, \textit{Propaganda and fake news} and about the debate on Twitter (\textit{Metadiscussion about Twitter vaccination discourse}) both make it into the top 20 topics (ranks 15 and 18, respectively). 
Topics 22 and 28 include heterogeneous collections of tweets: 22 using the \emph{SARSCoV2} term and hashtag (discussing a range of topics), 28 focusing on \emph{Statistics about vaccination uptake and policies}.  

The Mann-Kendall test reveals significant positive trends regarding the number of tweets for all of the top 30 topics, i.e., all of them gain increasing attention over time. 

\subsubsection{Themes}
As outlined above, many topics refer to similar issues with varying levels of granularity, e.g., the duty to vaccinate in general (17) vs. the duty to vaccinate at work (21). This restricts the informative value of the frequency rankings.  

Thus, we manually identify more general themes to investigate their salience over time and their relations to vaccination policy events. 
For this, we adopt the same workflow as for generating topic labels: each of the first two authors examines all topic labels and maps them to themes. 
On this basis, we arrive at the following final set of themes that include at least three topics: 
1) freedom and civic liberties, 2) safety and side effects of vaccinations, 3) effectiveness of vaccinations, 4) mobilization, 5) details about the vaccination campaign, 6) conspiracy theories, 7) country comparisons, 8) influential individuals and their stances or behaviors, 9) specific vaccines and 10) data about the pandemic. 

Tables \ref{tab:themes1} and \ref{tab:themes2} list the respective themes, the included topics, their summed-up frequencies, and average relative sentiment scores. 

\begin{table}[ht!]
    \centering
    \begin{tabularx}{\textwidth}{p{1.5cm}|l|l|X}

\textbf{Theme}&\textbf{$\#$}&\ensuremath \heartsuit&\textbf{Included topics}\\\hline
freedom and civic liberties&9975&-0.06&8 "Regulations", 11 "Lockdowns", 13 "Basic rights", 17 "Compulsory vaccination", 19 "Masks and mask mandate", 23 "Demonstrations and protests", 21 "Compulsory vaccinations at work", 34 "Compulsory vaccination II", 46 "Political parties on compulsory vaccination", 50 "Vaccination as solidarity", 60 "Vaccination as a personal free choice", 79 "Police and police officers", 82 "Travel", 84 "Nazis", 86 "Fascism, vaccination as suppression", 94 "Privileges for the vaccinated and enforcement of vaccinations", 107 "Compulsory vaccination compared to road safety", 112 "Concerts and musicians", 117 "Compulsory vaccination debate", 118 "Spahn's compulsory vaccination statement", 119 "Restaurant visits", 121 "Footballers and professional athletes", 125 "Discrimination of unvaccinated persons", 129 "Corona policies", 137 "Democracy, dictatorship, society", 144 "Compulsory vaccinations for all"\\\hline%

safety and side effects&5437&-0.17&16 "Vaccinations for elderly people", 25 "Vaccinations for children: medical views", 35 "Deaths due to or with Corona vaccination", 36 "Risks for pregnant women and infertility", 40 "Immediate vaccination side-effects", 48 "EMA", 49 "Side effects and risks with and without vaccination", 63 "Deaths after vaccination", 65 "Myocarditis risk after vaccination", 69 "AstraZeneca for specific age groups", 76 "Vaccination side effects", 83 "AstraZeneca vaccination stop", 91 "Safety of vaccinations", 104 "Effects of vaccinations on the menstrual cycle", 109 "Myocarditis risks", 110 "Pregnancy and risks", 127 "Side effects of vaccinations II", 134 "EMA recommendations and authorizations", 142 "Development of vaccines and their tests", 147 "Allergies and allergic reactions"\\\hline%

effectiveness&5040&-0.09&10 "Duration of vaccine protection", 12 "Mutations due to vaccinations, spread of the virus when vaccinated", 20 "Vaccine efficacy for Omicron", 27 "Immune system", 32 "vaccination protection", 98 "Vaccination protection and efficacy", 115 "Number of vaccinated people in hospitals", 116 "Infections after being vaccinated", 136 "Anecdotes of vaccinations and infections", 140 "Virus variants and mutations"\\\hline%

mobilization&4982&0.12&5 ""Do (not) get vaccinated"", 29 "\#AllesInDenArm", 41 "Appeals to get vaccinated", 47 ""I am vaccinated"", 54 ""I will not be vaccinated"", 68 "Opinion about own vaccination", 70 "Congratulations to being vaccinated", 71 "Communication of free vaccination appointments", 80 ""Got the second vaccination"", 85 "Booking of vaccination appointments", 89 "Disputations between vaccinated and unvaccinated persons", 97 "Vaccination appointments for children", 108 "Vaccination status updates", 124 "Personal reasons for or against getting vaccinated", 131 "Disputes", 141 "Booster shots", 143 "Calls to sign petitions"\\\hline%

vaccination campaign&2730&0.20&14 "Practical implementation", 33 "Vaccine purchase EU", 58 "Costs, monetary incentives and penalties", 59 "Priority groups", 64 "Bratwurst incentives", 66 "Who pays", 90 "Apps and digital vaccination certificates", 120 "Booked and free vaccination appointments"\\\hline%

conspiracy theories&1601&-0.18&15 "Propaganda and fake news", 37 "Bill Gates and vaccinations", 74 "Chips and implants"\\\hline%

\end{tabularx}
\caption{Mapping of topics to themes. $\#$ Number of tweets; \ensuremath \heartsuit average REL sentiment; \textit{Included topics} ranks of the topics regarding their frequencies in all tweets and their manually assigned labels}
\label{tab:themes1}
\end{table}

\begin{table}[ht!]
    \centering
    \begin{tabularx}{\textwidth}{p{1.5cm}|l|l|X}

\textbf{Theme}&\textbf{$\#$}&\ensuremath \heartsuit&\textbf{Included topics}\\\hline

country comparisons&6229&-0.13&4 "Situation in Germany (also in comparison with other countries)", 7 "Corona in Israel", 30 "Austria", 43 "Russia", 61 "Africa", 62 "Italy", 67 "France", 73 "Portugal", 75 "Vaccination of children in specific regions", 77 "China", 88 "Patent clearance", 92 "Great Britain", 105 "Israel", 106 "Global distribution of vaccines", 126 "Gibraltar", 138 "Switzerland"\\\hline%

influential individuals&4471&-0.08&3 "(prominent) vaccinated and unvaccinated men", 24 "Merkel", 31 "Lauterbach", 39 "Kimmich", 55 "Trump and Biden", 72 "Soeder", 96 "Sucharit Bhakdi", 111 "Kubicki", 128 "Politicians", 149 "BioNTech's founder"\\\hline%

specific vaccines&3276&0.01&9 "AstraZeneca vaccine", 26 "mRNA / gene-based vaccines", 42 "Sputnik vaccine", 56 "BioNTech", 81 "Novavax vaccine", 100 "Johnson $\&$ Johnson vaccine", 101 "Moderna vaccine", 130 "AstraZeneca vaccine II", 145 "J\&J vaccine"\\\hline%

data about the pandemic&754&0.08&44 "Statistics and headlines", 78 "Statistics about vaccination statuses", 113 "Statistics about the number of vaccinations", 133 "Statistics about vaccination rates in Germany"\\\hline%

\end{tabularx}
\caption{Mapping of topics to themes. $\#$ Number of tweets; \ensuremath \heartsuit average REL sentiment; \textit{Included topics} ranks of the topics regarding their frequencies in all tweets and their manually assigned labels}
\label{tab:themes2}
\end{table}

\textit{Freedom and civic liberties} contains tweets that discuss mandates and restrictions and debate force vs. free choice when it comes to Corona regulations. This theme has the highest number of tweets followed by \textit{country comparisons}, which bundles tweets comparing the situation in different countries. 
The third largest theme is \textit{safety and side effects}. This theme comprises tweets centering around the topic of short- and long-term side effects of vaccinations and risks connected to being vaccinated vs. unvaccinated. 
\textit{Effectiveness} designates tweets discussing the efficacy and usefulness of vaccinations, such as the duration of the offered protection or the virulence of vaccinated individuals. This theme also received high attention. 
In the \textit{mobilization} theme, we merge all topics containing messages motivating others to get vaccinated or to not get vaccinated. 
The stances, roles, or behaviors of authorities, leaders, or other influential individuals are discussed broadly as reflected by the sixth largest theme, \textit{influential individuals}. 
There is a high interest in discussing \textit{specific vaccines}. 
Details about the \textit{vaccination campaign} refer to different aspects of the vaccination strategy. This theme measures how policy events are reflected in the Twitter discourse. 
The theme \textit{conspiracy theories} contains tweets discussing different theories, e.g., concerning Bill Gates' motives regarding vaccinations, but also sarcastic tweets and tweets discussing news coverage and perceived propaganda on a meta-level. Therefore, many tweets belonging to this theme cannot be interpreted as a high level of belief in conspiracies or media distrust. Instead, it signals high attention to these topics. 
Twitter is also used to share \textit{data about the pandemic}, e.g., ratios of vaccinated persons or any other statistics. 

This analysis reveals that while many tweets concern health-related issues (safety and side effects, effectiveness, and specific vaccines), a very high number of tweets focus on the country level and the effects of policies on society. 
Overall, Twitter users seem to be similarly concerned about their freedom and civic liberties and health-related concerns. 

The Mann-Kendall test reveals that all themes exhibit a significant increasing trend regarding their tweet frequencies over the whole time span under investigation reflecting increased attention to vaccination-related discourse as a whole. 

\subsection{Topic and Theme Sentiments Over Time}\label{sec:rq3} 
To address RQ3, we next analyze the frequencies and sentiments of topics and themes over time. 

\subsubsection{Topic sentiments: complete time interval} 

The average sentiment scores support the analysis in Section \textit{Evolution of Vaccination Discourse in DACH Countries}. The sentiments for many topics are neither very negative nor positive when averaged over the whole time span, with a few exceptions. 
The topics with the most positive sentiments of the top 30 were \textit{Practical implementation} (0.39), \textit{Do (not) get vaccinated} (0.17), \textit{Basic rights} (0.14) and \textit{Duration of vaccination protection} (0.04). 
These are also the only topics with an overall positive sentiment. 
The \textit{Practical implementation} topic contains many tweets with people celebrating their vaccination appointments, \textit{Do (not) get vaccinated} many calls to action, and \textit{Basic rights} many references to positive concepts like freedom, privileges, and rights. The latter, however, contains many critical voices regarding vaccinations and compulsory vaccinations. 
The most negative of the top 30 topics were \textit{Propaganda and fake news} (-0.34), followed by \textit{Corona and other flu viruses} (-0.25), \textit{Corona in Israel} (-0.2), \textit{Immune system} (-0.19), \textit{Vaccine efficacy for Omicron} (-0.17) and \textit{Anecdotes: Experience with Corona vaccination} (-0.17). 
Also, all regulations and potential regulations (\textit{Regulations}, \textit{Lockdowns}, \textit{Compulsory vaccination}, \textit{Compulsory vaccination at work}, \textit{Masks}) are connected to negative sentiments. 
While sentiments are not to be interpreted as stances, i.e. negative sentiments do not necessarily signal disapproval, these negative scores suggest that these topics were coupled with a focus on negative aspects in the discussion. 

The Mann-Kendall test reveals significant negative trends regarding the relative sentiment for the following seven of the top 30 topics: 
\textit{(Prominent) vaccinated and unvaccinated men}, \textit{Vaccinations for elderly people}, \textit{Corona in Isreal}, \textit{Anecdotes: Experience with Corona vaccination}, \textit{Immune system}, \textit{Vaccine efficacy for Omicron}, \textit{Propaganda and fake news}. 

The only topics with significant positive trends are 
\textit{"Do (not) get vaccinated"} and \textit{Practical implementation}. 
No significant trends are found for the remaining topics. 

When analyzing only the relative positive respectively negative sentiment scores of all tweets on these topics, we observe a significant positive trend for positive sentiments and a significant negative trend for relative negative sentiments for all topics except \textit{Corona and other flu viruses}, for which there is no significant trend for the relative negative sentiments. 
This finding suggests that the discourse became more emotional and polarised over time. 
We will check next whether this also holds true for the vaccination discourse beyond the most prominent individual topics. 

\subsubsection{Theme sentiments: complete time interval} 
When analyzing only the relative positive respectively negative sentiment scores of all tweets, we observe a significant positive trend for positive sentiments and a significant negative trend for relative negative sentiments for all themes. While the impact on the consolidated sentiment varies across themes, this indicates that the vaccination discourse as a whole gets more emotional and polarised over time. 
The highest mean relative sentiment (i.e. the most positive sentiment) is connected to \textit{details of the vaccination campaign} (0.20) followed by \textit{mobilization} (0.12), which include the most positive individual topics, as outlined in the previous subsection. Together with \textit{data about the pandemic} (0.08) and \textit{specific vaccines} (0.01), these are the only themes with non-negative average sentiment scores. 
These are also the only themes with an overall positive trend. 
For all other themes, the negative trend is stronger than the positive one and they become significantly more negative. 
The lowest (i.e. most negative sentiment) is connected to \textit{conspiracy theories} (-0.18), \textit{safety and side effects} (-0.17), and \textit{country comparisons} (-0.13). 

We analyze the evolution of tweet frequencies for themes in more detail as part of the following section. 

\subsection{Relation to Phases and Policy Events} 
\begin{table}[ht!]
    \begin{minipage}{.5\linewidth}
    \begin{tabular}{p{0.2cm}|p{5cm}|l|l|l}
    \textbf{R}&\textbf{Topic}&\textbf{\#}&\textbf{\%}&\ensuremath\heartsuit\\\hline
    \multicolumn{5}{c}{\textbf{Sporadic cases}}\\
1&country comparisons&16&34.04\%&0.19\\\hline
2&effectiveness&6&12.77\%&0.33\\\hline
3&conspiracy theories&5&10.64\%&-0.20\\\hline
3&mobilization&5&10.64\%&0.20\\\hline
4&influential individuals&4&8.51\%&0.75\\\hline
5&details of the vaccination campaign&3&6.38\%&-1.33\\\hline
5&data about the pandemic&3&6.38\%&0.00\\\hline
5&freedom and civic liberties&3&6.38\%&0.00\\\hline
6&safety and side effects&1&2.13\%&0.00\\\hline
6&specific vaccines&1&2.13\%&0.00\\\hline

    \multicolumn{5}{c}{\textbf{Wave 1}}\\
1&conspiracy theories&176&24.14\%&0.06\\\hline
2&influential individuals&154&21.12\%&-0.14\\\hline
3&effectiveness&104&14.27\%&-0.05\\\hline
4&freedom and civic liberties&95&13.03\%&-0.01\\\hline
5&country comparisons&70&9.60\%&0.17\\\hline
6&mobilization&43&5.90\%&0.19\\\hline
7&data about the pandemic&39&5.35\%&-0.05\\\hline
8&safety and side effects&29&3.98\%&0.03\\\hline
9&details of the vaccination campaign&13&1.78\%&0.31\\\hline
10&specific vaccines&6&0.82\%&0.33\\\hline

    \multicolumn{5}{c}{\textbf{Summer Plateau 2020}}\\
1&country comparisons&202&24.16\%&-0.04\\\hline
2&freedom and civic liberties&137&16.39\%&0.05\\\hline
3&effectiveness&104&12.44\%&0.11\\\hline
4&conspiracy theories&92&11.00\%&-0.02\\\hline
5&influential individuals&91&10.89\%&0.09\\\hline
6&mobilization&62&7.42\%&0.35\\\hline
7&specific vaccines&50&5.98\%&-0.08\\\hline
8&safety and side effects&42&5.02\%&-0.17\\\hline
9&details of the vaccination campaign&30&3.59\%&0.13\\\hline
10&data about the pandemic&26&3.11\%&0.08\\\hline

    \multicolumn{5}{c}{\textbf{Wave 2}}\\
1&freedom and civic liberties&1139&17.27\%&-0.08\\\hline
2&country comparisons&1101&16.69\%&-0.06\\\hline
3&influential individuals&865&13.12\%&0.06\\\hline
4&safety and side effects&812&12.31\%&-0.18\\\hline
5&specific vaccines&738&11.19\%&-0.05\\\hline
6&effectiveness&610&9.25\%&-0.10\\\hline
7&mobilization&532&8.07\%&0.16\\\hline
8&details of the vaccination campaign&403&6.11\%&0.06\\\hline
9&conspiracy theories&266&4.03\%&-0.10\\\hline
10&data about the pandemic&129&1.96\%&0.06\\\hline

           \end{tabular}
     \end{minipage}
     \caption{Theme frequency, all tweets (different phases of the pandemic). \textit{R} Rank; \textit{\#} Number of tweets; \textit{\%} Relative number of tweets; \ensuremath\heartsuit Average sentiment per tweet (REL)}
    \label{tab:themesOverTime1}
\end{table}
     
\begin{table}[ht!]     
     \begin{minipage}{.5\linewidth}
     \begin{tabular}{p{0.2cm}|p{5cm}|l|l|l}
    \textbf{R}&\textbf{Topic}&\textbf{\#}&\textbf{\%}&\ensuremath\heartsuit\\\hline
    \multicolumn{5}{c}{\textbf{Wave 3}}\\
1&safety and side effects&1611&17.55\%&-0.14\\\hline
2&freedom and civic liberties&1390&15.14\%&-0.04\\\hline
3&specific vaccines&1384&15.08\%&-0.01\\\hline
4&country comparisons&1254&13.66\%&-0.06\\\hline
5&mobilization&995&10.84\%&0.19\\\hline
6&influential individuals&789&8.60\%&-0.02\\\hline
7&effectiveness&711&7.75\%&-0.09\\\hline
8&details of the vaccination campaign&687&7.48\%&0.12\\\hline
9&conspiracy theories&220&2.40\%&-0.10\\\hline
10&data about the pandemic&138&1.50\%&0.03\\\hline

 \multicolumn{5}{c}{\textbf{Summer Plateau 2021}}\\
1&freedom and civic liberties&742&18.10\%&-0.09\\\hline
2&country comparisons&675&16.46\%&-0.26\\\hline
3&mobilization&569&13.88\%&0.14\\\hline
4&details of the vaccination campaign&504&12.29\%&0.32\\\hline
5&safety and side effects&487&11.88\%&-0.24\\\hline
6&effectiveness&359&8.76\%&-0.17\\\hline
7&influential individuals&324&7.90\%&-0.10\\\hline
8&specific vaccines&234&5.71\%&-0.03\\\hline
9&conspiracy theories&130&3.17\%&-0.23\\\hline
10&data about the pandemic&76&1.85\%&0.17\\\hline

    \multicolumn{5}{c}{\textbf{Wave 4}}\\
1&freedom and civic liberties&5140&27.77\%&-0.07\\\hline
2&effectiveness&2362&12.76\%&-0.09\\\hline
3&country comparisons&2341&12.65\%&-0.20\\\hline
4&mobilization&2335&12.61\%&0.07\\\hline
5&safety and side effects&2073&11.20\%&-0.14\\\hline
6&influential individuals&1756&9.49\%&-0.16\\\hline
7&details of the vaccination campaign&958&5.18\%&0.20\\\hline
8&specific vaccines&702&3.79\%&0.13\\\hline
9&conspiracy theories&543&2.93\%&-0.27\\\hline
10&data about the pandemic&301&1.63\%&0.13\\\hline

    \multicolumn{5}{c}{\textbf{Wave 5}}\\
1&freedom and civic liberties&1403&28.84\%&-0.06\\\hline
2&effectiveness&831&17.08\%&-0.09\\\hline
3&country comparisons&622&12.79\%&-0.04\\\hline
4&influential individuals&513&10.54\%&-0.11\\\hline
5&mobilization&483&9.93\%&0.10\\\hline
6&safety and side effects&428&8.80\%&-0.30\\\hline
7&specific vaccines&189&3.88\%&0.04\\\hline
8&conspiracy theories&180&3.70\%&-0.39\\\hline
9&details of the vaccination campaign&167&3.43\%&0.46\\\hline
10&data about the pandemic&49&1.01\%&-0.02\\\hline
     \end{tabular}
     \end{minipage}

    \caption{Theme frequency, all tweets (different phases of the pandemic). \textit{R} Rank; \textit{\#} Number of tweets; \textit{\%} Relative number of tweets; \ensuremath\heartsuit Average sentiment per tweet (REL)}
    \label{tab:themesOverTime2}
\end{table}

Next, we analyze the attention to themes during different pandemic phases and in relation to policy events to answer RQ4. 

\subsubsection{Relation to pandemic phases} 
Tables \ref{tab:themesOverTime1} and \ref{tab:themesOverTime2} list the frequencies and connected sentiments of all themes during different phases of the pandemic. Since the phases differ in their durations and the number of tweets has been increasing overall over time, we also list the relative number of tweets as a percentage of tweets belonging to any of the themes. 

We exclude the phase \textit{Sporadic cases} from the following analysis as it did not contain enough tweets to derive meaningful rankings. 

The themes \textit{freedom and civic liberties} and \textit{country comparisons} were prominent throughout all phases of the pandemic: they have the highest (i.e. the top) mean ranks across all time intervals (1.71 and 2.86, respectively) and a high rank stability with standard deviations of 1.11 and 1.35, respectively. 
For \textit{freedom and civic liberties}, we observe an increased frequency, both absolute and relative, in the last two phases, i.e. it received more attention during the later phases than during the early ones. 
Both themes have their lowest rank during the first wave and their second lowest during Wave 3. 
The top 3 themes \textit{effectiveness} (mean rank 4.14) also has its lowest rank in Wave 3, i.e. there it received its least attention.  
This phase was dominated by the \textit{safety and side effects} theme (mean rank 5.29). Ranking third in Wave 3, behind \textit{freedom and civic liberties}, we find \textit{specific vaccines}. In other phases, \textit{specific vaccines} received attention, too, but to a lesser degree (mean rank 6.86). 
\textit{Safety and side effects}, \textit{specific vaccines} and \textit{effectiveness} show the greatest fluctuations in ranks with 2.43, 2.27, and 2.12 standard deviation, respectively. Only \textit{conspiracy theories} fluctuates more (3.21). 

\begin{figure}[ht]
    \centering
    \includegraphics[width=\textwidth]{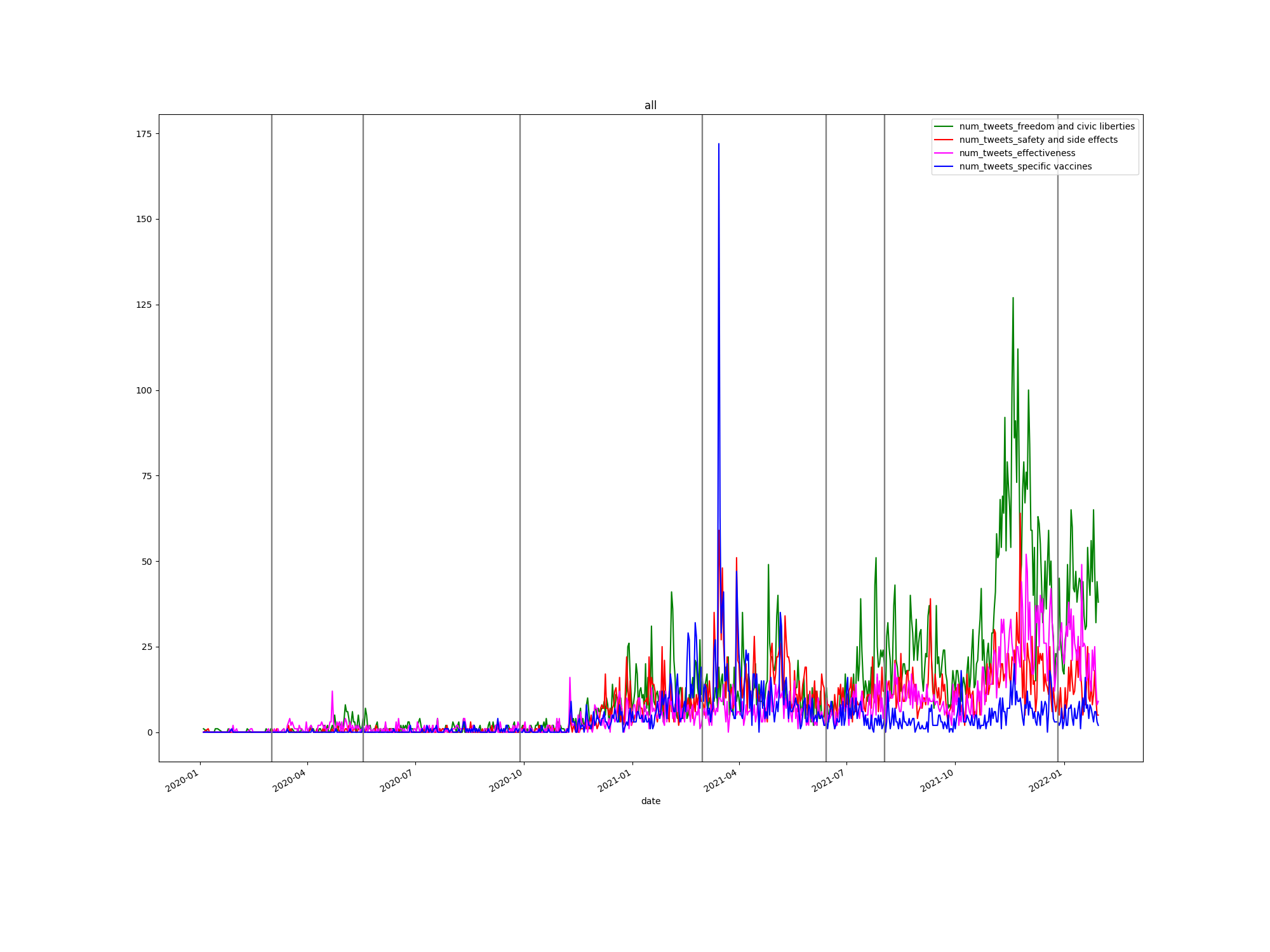}
    \caption{Tweet frequencies of the themes \textit{Freedom and civic liberties} (green), \textit{Safety and side effects} (red), \textit{Effectiveness} (magenta) and \textit{Specific vaccines} (blue). Grey lines mark the start of a pandemic phase}
    \label{fig:selectedThemes}
\end{figure}

The development of the highly fluctuating themes \textit{Safety and side effects}, \textit{Effectiveness}, \textit{Specific vaccines} and the dominating theme \textit{Freedom and civic liberties} is illustrated in Figure \ref{fig:selectedThemes}. 
This analysis shows that attention to topics in Wave 3 differs from the other phases and that directly vaccine-related themes were the most unstable. We will investigate possible reasons in more detail in section \ref{sec:eventInterpretations}. 

\subsubsection{Relation to policy events} \label{sec:eventInterpretations} 

\begin{figure}[ht]
    \centering
    \includegraphics[width=\textwidth]{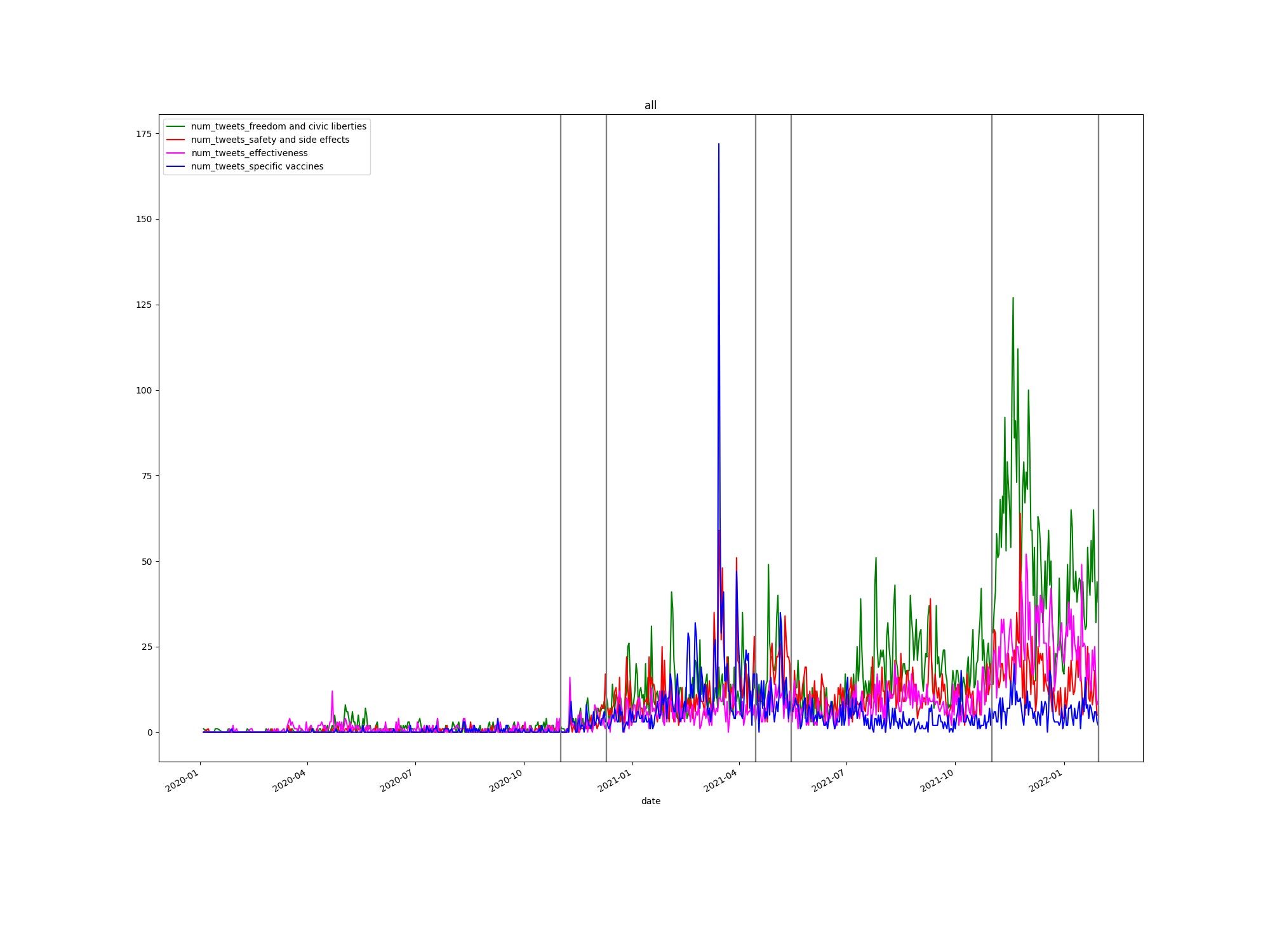}
    \caption{The selected themes over different policy phases. Grey lines mark the beginning of a policy phase}
    \label{fig:selectedThemesPolicyPhases}
\end{figure}

In order to reveal possible connections to policy actions, we investigate the relation of change points and peaks for the most fluctuating and dominant themes in \ref{fig:selectedThemes}. 
First, we relate them to the policy phases introduced in Table \ref{tab:policyPhases} with Figure \ref{fig:selectedThemesPolicyPhases}. 

Compared to Figure \ref{fig:selectedThemes}, the development of the themes seems to align better with the policy phases than with the pandemic phases as classified by the RKI. 

\begin{figure}[ht]
    \centering
    \includegraphics[width=\textwidth]{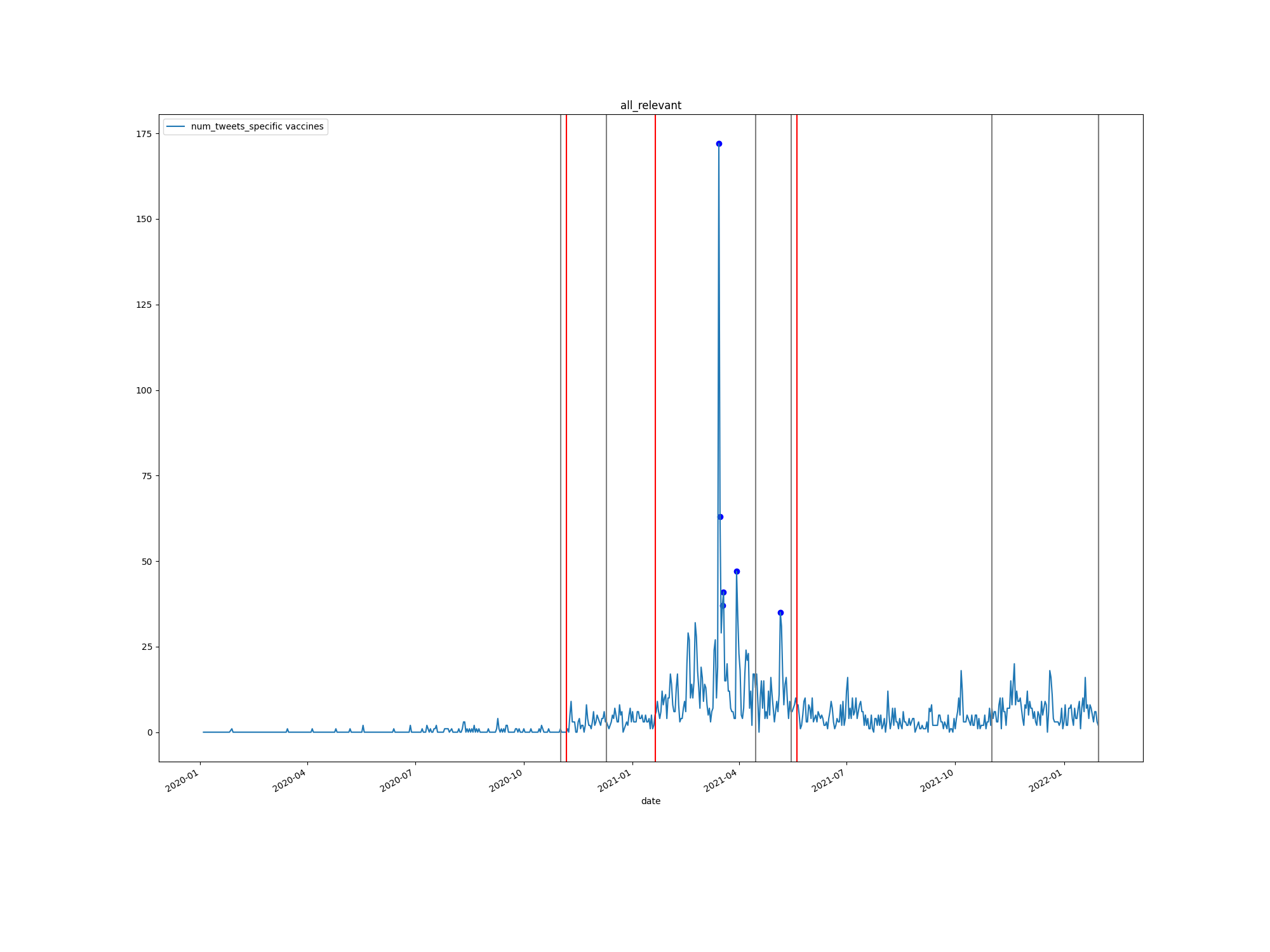}
    \caption{Change points and peaks in tweet frequency for the \textit{Specific vaccines} theme. Grey lines mark the policy phases, red lines the change points, blue dots peaks. Change points occur at 06/11/2020, 20/01/2021, 20/05/2021. Peaks are detected at 15/03/2021, 16/03/2021, 18/03/2021, 19/03/2021, 30/03/2021, 06/05/2021.}
    \label{fig:events_vaccines}
\end{figure}

\begin{figure}[ht]
    \centering
    \includegraphics[width=\textwidth]{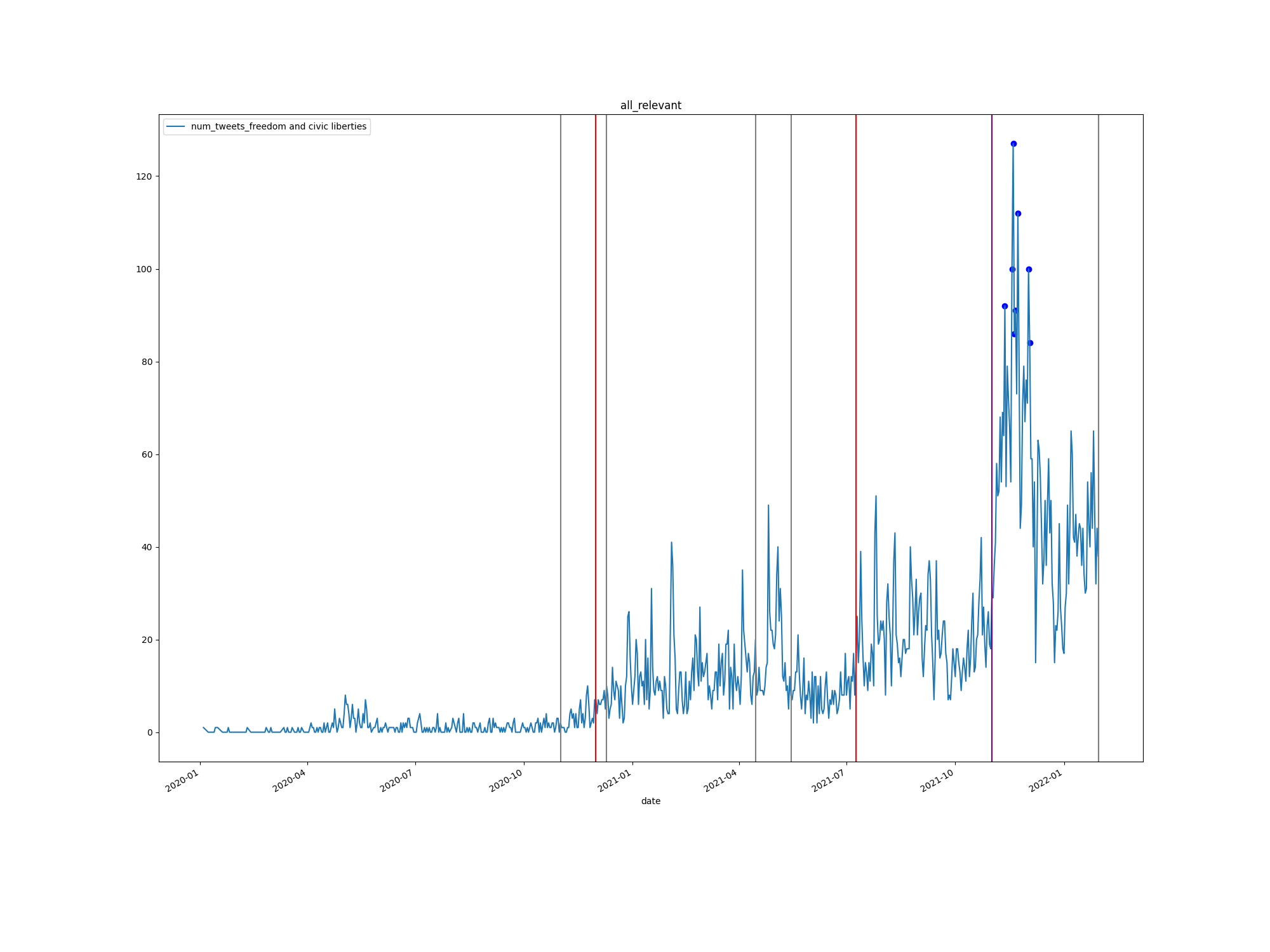}
    \caption{Change points and peaks in tweet frequency for the \textit{Freedom and civic liberties} theme. Grey lines mark the policy phases, red lines the change points, purple lines policy phases and change points at the exact same day, blue dots peaks. Change points occur at 01/12/2020, 09/07/2021, 01/11/2021. Peaks are detected at 12/11/2021, 18/11/2021, 19/11/2021, 20/11/2021, 21/11/2021, 23/11/2021, 02/12/2021, 03/12/2021. }
    \label{fig:events_freedom}
\end{figure}

\begin{figure}[ht]
    \centering
    \includegraphics[width=\textwidth]{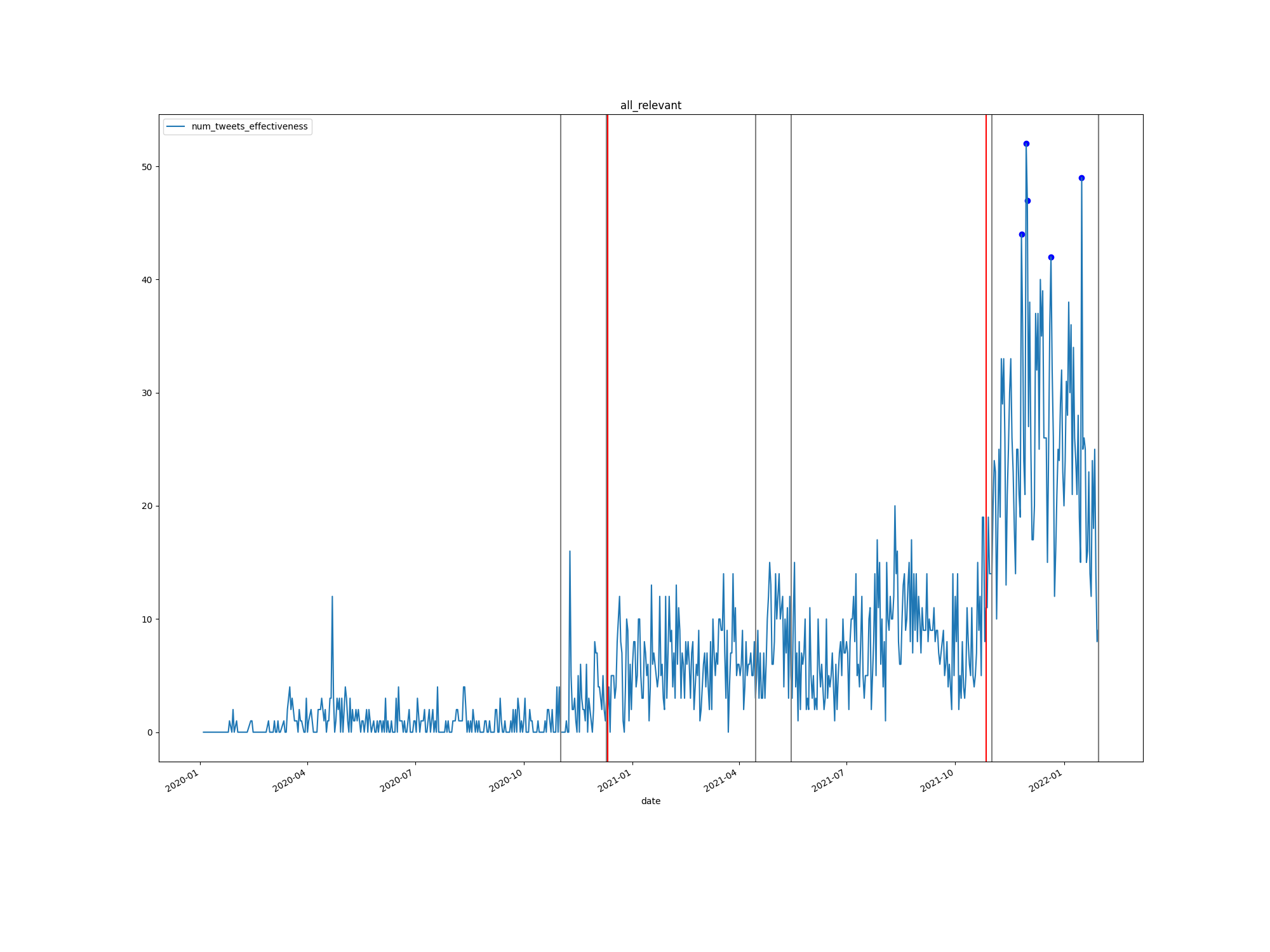}
    \caption{Change points and peaks in tweet frequency for the \textit{Effectiveness} theme. Grey lines mark the policy phases, red lines the change points, blue dots peaks. Change points occur at 11/12/2020, 27/10/2021. Peaks are detected at 26/11/2021, 30/11/2021, 01/12/2021, 21/12/2021, 16/01/2022.}
    \label{fig:events_effectiveness}
\end{figure}

\begin{figure}[ht]
    \centering
    \includegraphics[width=\textwidth]{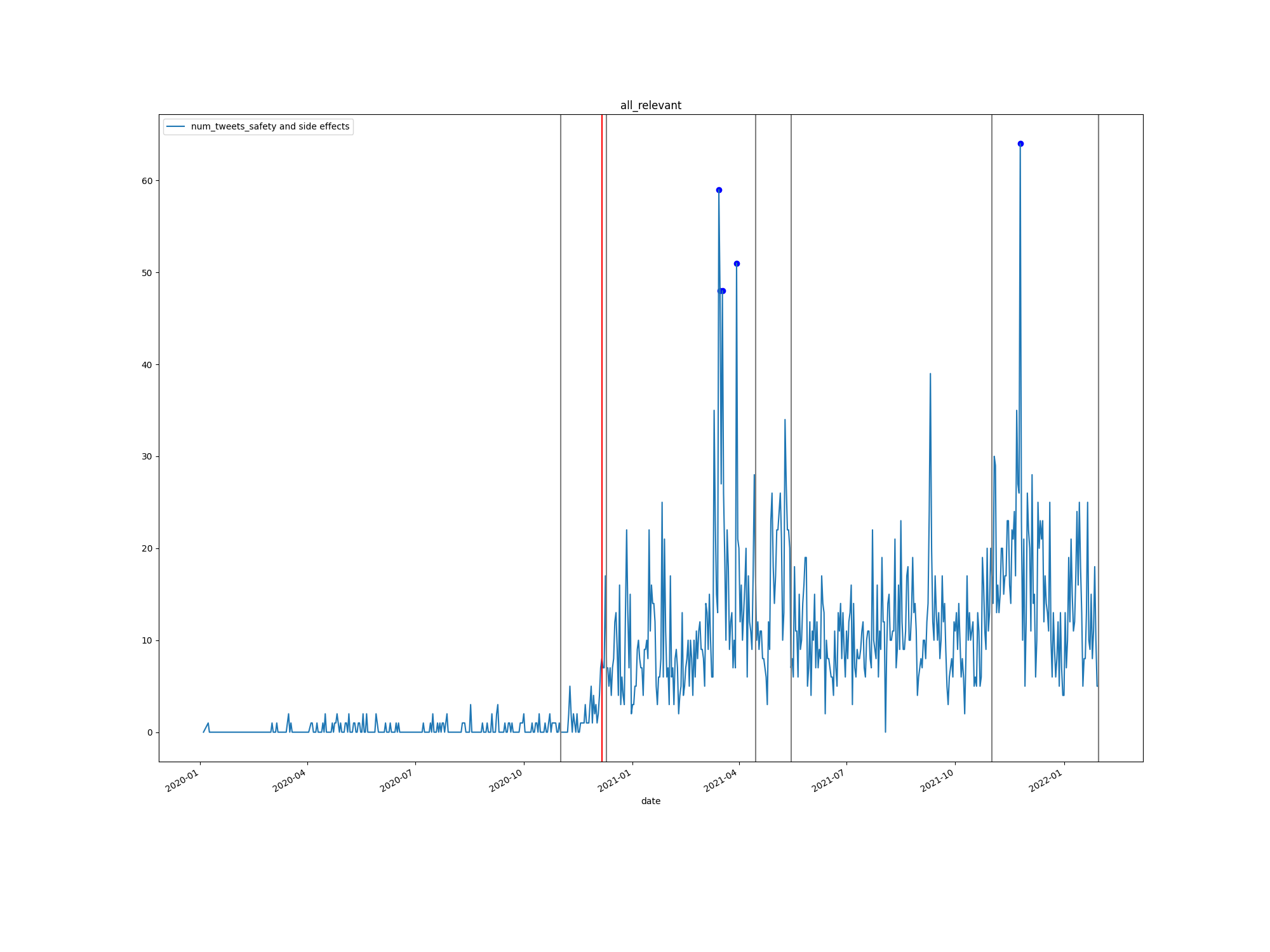}
    \caption{Change points and peaks in tweet frequency for the \textit{Safety and side effects} theme. Grey lines mark the policy phases, red lines the change points, blue dots peaks. Change points occur at 06/12/2020. Peaks are detected at 15/03/202, 16/03/2021, 18/03/2021, 30/03/2021}
    \label{fig:events_safety}
\end{figure}

This is further supported by the analysis of change points and peaks, as visualized in Figures \ref{fig:events_vaccines}, \ref{fig:events_freedom}, \ref{fig:events_effectiveness}, \ref{fig:events_safety}. 

Considering the policy phases outlined in Table \ref{tab:policyPhases}, phase I from November until mid-December 2020 marks the beginning of official COVID-19 vaccination policies in the DACH region. This period went along with change points in tweet numbers for \textit{Specific vaccines}, which started to be discussed frequently after this point. 
 
In phase II starting mid-December 2020, the first vaccination recommendations and strategies for COVID vaccinations were published and the first vaccines were authorized before vaccinations finally started on December 23 (Switzerland) and 27 (Austria and Germany). 
This went along with changes in tweet frequencies according to the change point analysis for \textit{Effectiveness}, \textit{Safety and side effects}, and \textit{Freedom and civic liberties} which received increasing attention from then on. 
In the same phase on March 15, 2021, AstraZeneca vaccinations were halted in Germany due to safety concerns (cf. Table \ref{tab:eventsD}) before they were resumed on March 25. 
We observe peaks for the themes \textit{Safety and side effects} and \textit{Specific vaccines} at the exact same time. 

On May 6, 2021, during the third phase, vaccinations with the unpopular AstraZeneca vaccine were possible for all individuals, regardless of priority group membership. A peak for \textit{Specific vaccines} can be found on the same day. 

Policy phase IV contained few policy events and few topic rank fluctuations, change points, and peaks. \textit{Freedom and civic liberties} was the dominant theme. 
Restrictions for unvaccinated persons were discussed and finally implemented in Germany in August 2021. 

November 2021 marked the month of booster recommendations and authorizations of vaccines for children. Also, on November 15, 2021, there was a nationwide lockdown for unvaccinated persons in Austria. 
This and the following month contained the all-time peaks of the \textit{Freedom and civic liberties} and the \textit{Effectiveness} themes. 

This analysis suggests that the high fluctuation of attention to the themes \textit{Safety and side effects}, \textit{Effectiveness}, and \textit{Specific vaccines} were related to specific policy actions. 

While vaccine safety and individual vaccines had been in public attention from the start, the decision to halt AstraZeneca vaccinations in Germany seemed to have had a big impact on the focus of public attention, fueling the controversy. However, soon other topics were again discussed more broadly. 
The theme of freedom and civic liberties received more attention early on and only lost its dominating position shortly during the peak of the AstraZeneca debate. Shortly after, it resumed being the focus of attention. Especially during the later phases of the pandemic, also restrictions for unvaccinated persons were debated, this theme was debated strongly. 
The recommendation to receive booster shots and the debates about the recommended time between the second and third vaccination seemed to have sparked a contested debate about the overall vaccine effectiveness. Concerns about safety and side effects also resurfaced at that time. 

Overall, the debate on Twitter evolved strongly around the question of civic liberties. 
The public opinion about vaccinations in the DACH region did not seem to have become more positive or the concerns fewer, yet the willingness to get vaccinated increased. 
This analysis suggests that this can most likely be attributed to the incentives to get vaccinated or possible restrictions when not getting vaccinated. 
Another possible interpretation is that the discussion of these topics might have contributed to removing the medical concerns from the focus of attention, thereby removing an important driver of vaccine hesitancy: as \cite{solisArceEtAl2021} find in their survey-based study for low- and middle-income countries:
"Vaccine acceptance is explained mainly by an interest in personal protection against COVID-19, whereas concerns about side effects are the most common reasons for hesitancy." 

However, the increasing polarization of the debate suggests that this came at the price of deepening the rifts between proponents and opponents of vaccinations, further politicizing the issue. 

\section{Summary and Discussion} 
Our empirical study used NLP methods to detect and analyze 199,207 tweets about COVID-19 vaccinations in the DACH region (Austria, Germany, Switzerland). The results reflect that the topic was controversially discussed: we find that the total number of tweets about this important societal issue increased over time, and the sentiments in the discourse became both more polarized and more negative (RQ1), cf. Figure \ref{fig:summedUpSentimentsOfTweets}. Generally, discourse about COVID-19 vaccinations has been significantly more negative than the average discourse on Twitter during the same time period (Figure \ref{fig:sentiments}).

Investigating RQ2 and RQ3, we find that the Twitter discourse data reveal fluctuations in the topics and themes (cf. \ref{tab:themes}) that are at the center of public attention: while medical concerns such as the safety and side-effects of vaccinations were prominently discussed early in the debate and concerning specific events (cf. Tables \ref{tab:themesOverTime1} and \ref{tab:themesOverTime2}, Figures \ref{fig:selectedThemesPolicyPhases}, \ref{fig:events_vaccines},  \ref{fig:events_effectiveness}, \ref{fig:events_safety}), the focus increasingly shifted to a discussion of broader societal concerns: especially those regarding freedom and civic liberties (cf. Tables \ref{tab:themesOverTime1} and \ref{tab:themesOverTime2}, Figures \ref{fig:selectedThemesPolicyPhases} and \ref{fig:events_freedom}). 
At the same time, vaccination acceptance and uptake were low early in the pandemic and increased over time (\cite{DessonEtAl2020}, Table \ref{tab:rkiPhases}). 
Our investigations into RQ4 give insights into possible drivers of these changes: Figures \ref{fig:selectedThemes} and \ref{fig:selectedThemesPolicyPhases} show that shifts in the discourse align better with policy phases than pandemic phases. Figures \ref{fig:events_vaccines}, \ref{fig:events_freedom}, \ref{fig:events_effectiveness} and \ref{fig:events_safety} illustrate that attention peaks to themes were related to policy events such as halting vaccinations with AstraZeneca or incentivizing vaccinated persons. 
Thus, policies implementing incentives to get vaccinated or restrictions for unvaccinated people seem to have been successful in increasing vaccination uptake, either by the incentives/restrictions themselves or by removing the focus of public attention from medical concerns. 
However, based on our findings in RQ3, these policies did not increase citizens' positive sentiments about vaccinations in general (cf. Figure \ref{fig:relSentimentOfTweets}, Tables \ref{tab:themesOverTime1} and \ref{tab:themesOverTime2}) but rather increased polarization (cf. page \pageref{sec:rq3}). 

Moreover, these findings suggest that information campaigns about medical concerns might have been helpful in addressing citizens' concerns during the early stages of the pandemic. These concerns might have never been cleared for many COVID-19 vaccination critics. Instead, debates about compulsory vaccinations and benefits for vaccinated people sparked a discussion about freedom and civic liberties, which received more attention than medical concerns. This demonstrates that insights into the different reasons for vaccine hesitancy among the citizenry are crucial to design and implement adequate policy responses. While information campaigns about side-effects may be effective in addressing medical concerns, such campaigns would hardly convince people who oppose vaccinations because they feel that their individual freedom is being violated. 

It is in line with this interpretation that vaccine hesitancy in the DACH region decreased compared to other European countries, but sentiments did not become more positive (RQ1). 
The situation in other countries received considerable attention during all phases of the pandemic (RQ3, cf. Tables \ref{tab:themesOverTime1} and \ref{tab:themesOverTime2}). This suggests that policies and debates in other countries may strongly influence citizens' opinions and behaviors. Citizens seek orientation beyond the borders of their country. This finding indicates the need for international solutions and cooperation. 

Lastly, our findings suggest that analyzing online discourse data can yield valuable insights for policy-makers regarding topics of interest and attention to public concerns in highly dynamic contexts such as the COVID-19 pandemic. Online discourses can be a fruitful data source in addition to traditional survey data. 
The findings presented in this study contain relevant information about the possible relationship between policy events and public opinion that could inform political decision-makers. Our analyses suggest that the changes in public attention align better with different policy phases than with phases reflecting the infection rates alone. 
Yet, our analysis has several limitations. 
First, Twitter users are not representative of the whole population. 
Therefore, analyzing tweets can serve to analyze fluctuations and tendencies but should not be interpreted as a representation of general public opinion. 
Second, while we tried to ingest as little prior knowledge into our analyses as possible, opting for a primarily data-driven approach, our analysis is influenced by the choice of policy events and the segmentation into pandemic and policy phases. We did not investigate other events beyond infection rates and policies that may influence or relate to the discourse, such as news or social media discussions. 
Third, the assignment of themes was based on automatically generated topics but is still subjective. Different abstraction levels would have also been valid. 
The same applies to the generation of topics as such. The generated topics are not entirely selective; e.g., topics in the cluster "individual vaccines", e.g., the AstraZeneca topic, contain tweets that also discuss side-effects and vice versa. The same is true for the topic clusters discussing the effectiveness of vaccinations and the Omicron variant. To not produce too much noise, we decided to assign each tweet to the most probable cluster and not assign any cluster for low-confidence assignments. For future work, we will investigate the effects of assigning tweets to multiple clusters, controlling for the noise generated by different thresholds and parameters, and assessing topic cluster stability in different settings. 
Last, our generated data can be analyzed further to draw more detailed insights on additional topics relating to the formation and change of public opinion related to COVID-19 vaccinations. For example, while the attitudes and behaviors of influential individuals appeared to play an essential role in the public discourse on Twitter, it would be interesting to differentiate between different types of individuals, such as politicians or celebrities, advocates and opponents of vaccinations, and genuine vs. false information in their statements to gain more insights on the role of issues of trust and misinformation.

\bibliographystyle{abbrv} 
\bibliography{bib} 

\end{document}